\documentclass{aastex631}

\usepackage{enumerate}

\submitjournal{ApJ}

\shorttitle{Observing SN Neutrinos: II. Impact of EOS}
\shortauthors{Nakazato et al.}
\graphicspath{{./}{figures/}}

\begin{document}

\title{Observing Supernova Neutrino Light Curves with Super-Kamiokande:\\ II. Impact of the Nuclear Equation of State}

\correspondingauthor{Ken'ichiro Nakazato}
\email{nakazato@artsci.kyushu-u.ac.jp}

\author[0000-0001-6330-1685]{Ken'ichiro Nakazato}
\affiliation{Faculty of Arts and Science, Kyushu University, Fukuoka 819-0395, Japan}

\author{Fumi Nakanishi}
\affiliation{Department of Physics, Okayama University, Okayama 700-8530, Japan}

\author{Masayuki Harada}
\affiliation{Department of Physics, Okayama University, Okayama 700-8530, Japan}

\author[0000-0003-0437-8505]{Yusuke Koshio}
\affiliation{Department of Physics, Okayama University, Okayama 700-8530, Japan}
\affiliation{Kavli Institute for the Physics and Mathematics of the Universe (Kavli IPMU, WPI), Todai Institutes for Advanced Study, \\ The University of Tokyo, Kashiwa 277-8583, Japan}

\author[0000-0002-7443-2215]{Yudai Suwa}
\affiliation{Department of Earth Science and Astronomy, The University of Tokyo, Tokyo 153-8902, Japan}
\affiliation{Center for Gravitational Physics, Yukawa Institute for Theoretical Physics, Kyoto University, Kyoto 606-8502, Japan}
\affiliation{Department of Astrophysics and Atmospheric Sciences, Kyoto Sangyo University, Kyoto 603-8555, Japan}

\author[0000-0002-9224-9449]{Kohsuke Sumiyoshi}
\affiliation{National Institute of Technology, Numazu College of Technology, Numazu 410-8501, Japan}

\author[0000-0003-1409-0695]{Akira Harada}
\affiliation{Institute for Cosmic Ray Research, The University of Tokyo, Kashiwa 277-8582, Japan}
\affiliation{Interdisciplinary Theoretical and Mathematical Sciences Program (iTHEMS), RIKEN, Wako, Saitama 351-0198, Japan}

\author{Masamitsu Mori}
\affiliation{Department of Physics, Kyoto University, Kyoto 606-8502, Japan}
\affiliation{Department of Earth Science and Astronomy, The University of Tokyo, Tokyo 153-8902, Japan}

\author{Roger A. Wendell}
\affiliation{Department of Physics, Kyoto University, Kyoto 606-8502, Japan}
\affiliation{Kavli Institute for the Physics and Mathematics of the Universe (Kavli IPMU, WPI), Todai Institutes for Advanced Study, \\ The University of Tokyo, Kashiwa 277-8583, Japan}



\begin{abstract}
The late-time evolution of the neutrino event rate from supernovae is evaluated for Super-Kamiokande  using simulated results of proto--neutron star (PNS) cooling. In the present work we extend the result of \citet{2019ApJ...881..139S}, which studied the dependence on the PNS mass, but focus on the impact of the nuclear equation of state (EOS).
We find that the neutrino event rate depends on both the high-density and low-density EOS, where the former determines the radius of the PNS and the latter affects its surface temperature.
Based on the present evaluation of the neutrino event rate we propose a new analysis method to extract the time variability of the neutrino average energy taking into account the statistical error in the observation.
\end{abstract}

\keywords{Supernova neutrinos --- Neutrino astronomy --- Neutrino telescopes --- Nuclear astrophysics --- Neutron stars --- Core-collapse supernovae}


\section{Introduction} \label{sec:intro}

The observation of a Galactic supernova has been long-awaited by both astronomers and physicists. 
Indeed, core-collapse supernovae provide a high-energy environment that is otherwise inaccessible to terrestrial experiments and whose neutrino emission is driven by extreme conditions realized within the supernova core.
When the stellar core is compressed beyond the nuclear density ($\sim\,3\times10^{14}$~g~cm$^{-3}$) during the collapse, its  temperature rises to several tens of MeV. 
This hot supernova core (proto--neutron star, PNS) subsequently cools via neutrino emission leaving behind a cold neutron-rich neutron star.  
Observation of the neutrinos emitted during the cooling provides a probe of the nuclear equation of state (EOS) that covers the relevant temperature and proton-fraction ranges. Super-Kamiokande is expected to be a powerful tool for this purpose.\footnote{There are many other instruments that will also detect significant numbers of neutrinos  \citep[see][for instance]{scho18, 2020arXiv201100035A}.}

The nuclear EOS is essential to predict the total energy of neutrinos emitted from the supernova explosion since it determines the binding energy of the neutron star that results from the massive star. 
This binding energy is then converted to thermal energy of the central object, producing an abundance of neutrinos via several processes. 
Indeed, most of the energy emitted from supernovae is carried away by these neutrinos.
From the observation of SN 1987A \citep{1987PhRvL..58.1490H,1987PhRvL..58.1494B,1988PhLB..205..209A}, the total estimate for energy carried by the supernova neutrinos was confirmed to be consistent with a range of the binding energies from neutron stars \citep{1987PhLB..196..267S,1989ApJ...340..426L}.
The precise binding energy is determined by the still uncertain EOS through the gravitational mass and radius of the neutron star. 
If the dense matter at the supernova core is easily compressible (referred to as ‘soft’), the neutron star's density becomes large and its radius becomes small. 
As a result, the total energy of emitted neutrinos increases due to the large binding energy of neutron star. 
On the other hand, if the matter is less compressible at high densities (referred to as ‘stiff’), the neutron star's radius is large, its binding energy small, and the total energy of neutrinos is also reduced. 
For these reasons, the general behavior of the EOS is crucial to understanding supernova neutrinos. 

In addition to the energetics described above, the details of the nuclear EOS are important to determine the neutrino signal from the PNS.
Though neutrinos of all flavors are produced and trapped inside the high density matter of the collapsing star's core they gradually diffuse out as the PNS forms, carrying away energy and cooling it in the process. 
This emission accordingly reflects the environment of the stellar matter, namely its density, temperature, and composition.
In a first approximation, the neutrino energies are proportional to the temperature and their fluxes are determined by the matter density gradient. 
More precisely, the neutrino distributions are determined by the density and temperature at thermal and chemical equilibrium, but the neutrinos are transported outside through non-equilibrium neutrino processes.
A future detection of supernova neutrinos with large detectors would therefore provide important clues as to the properties of the hot and dense nuclear matter in the dying star.

While considerable effort has been devoted to modeling supernova dynamics in its early stage (up to $\sim$2~s after the bounce) in order to elucidate the explosion mechanism \citep[see][for neutrinos from supernovae and its relation with hydrodynamics]{2017hsn..book.1575J,2019ARNPS..69..253M}, the neutrino emission during the late phase is also important to understand the supernova energetics and the compact object formation. 
Pioneering work in this direction was done by \citet{1998ApJ...496..216T}, which  studied the detection statistics for Super-Kamiokande using Monte Carlo (MC) simulations based on a single model of long-term supernova neutrino emission, known as the Wilson model \citep{beth85}.
Recently, some long-term simulations have been done with modern hydrodynamical evolution \citep{2010PhRvL.104y1101H,2012PhRvD..85h3003F,2014PASJ...66L...1S,2021PhRvD.103b3016L,2020arXiv201016254M,2021arXiv210211283N} and quasi-stationary evolution \citep{1995A&A...296..145K,2013ApJS..205....2N,2017PhRvD..96d3015C} techniques. 
To predict neutrino observations from the next Galactic supernova, detailed knowledge of hot and dense matter as well as the relevant uncertainties needs to be included in sophisticated numerical simulations.
These details may affect the neutrino signal by way of influencing the properties of the PNS \citep{1988ApJ...334..891B,1995A&A...303..475S,1999ApJ...513..780P,2012PhRvL.108f1103R,2018PhRvC..97c5804N,2019ApJ...887..110S,2019ApJ...878...25N}.

In \citet[hereafter Paper I]{2019ApJ...881..139S}, we systematically performed PNS cooling calculations over 10 seconds by changing both the PNS mass from $\sim$1.2 $M_\odot$ to 2.05 $M_\odot$ and the initial entropy profile taking into account the explosion details and the early PNS thermodynamic profile. 
Based on these calculations we evaluated the supernova neutrino signals expected in Super-Kamiokande with a focus on the last observable event.
We demonstrated that the duration of the neutrino observation strongly depends on the mass of the PNS born in the supernova. Furthermore, we proposed a novel backward-in-time analysis focused on the late phase of the neutrino light curve to extract properties of the PNS independent of the early-time evolution.
However, the EOS dependence of the neutrino light curve was not discussed in Paper I. In this paper, we investigate the impact of the nuclear EOS on this backward time analysis.

The late-time behavior of the supernova neutrino burst is sensitive to the nuclear EOS.
\citet{2019ApJ...878...25N,2020ApJ...891..156N} systematically simulated the PNS cooling process using parameterized EOS models.
Finding that the neutrino light curve is characterized by the gravitational mass, $M_{\rm NS,g}$, and radius, $R_{\rm NS}$, of the resulting neutron star,
they proposed a method to estimate these parameters.
Since the binding energy of the neutron star determines the total energy of emitted neutrinos and depends on $M_{\rm NS,g}$ and $R_{\rm NS}$ \citep{2001ApJ...550..426L},
properties of the nascent neutron star can be extracted from neutrino observations.
Note that the mass--radius relation of neutron stars is mainly determined by the EOS at supranuclear densities.
On the other hand, the EOS at subnuclear densities affects the average energy of emitted neutrinos because heavy nuclei
residing in this regime efficiently interact with neutrinos via coherent effects~\citep{2018PhRvC..97c5804N}.
Accordingly, the average neutrino energy is higher for an EOS with a larger mass number of heavy nuclei.

This paper presents the first attempt to include the EOS dependence of the late-phase supernova neutrino event rate.
In \S\ref{sec:eos} we briefly review the nuclear EOS and introduce the models adopted in the present study.
Neutrino emission from the cooling PNS is described in \S\ref{sec:pns} and a procedure to estimate the neutrino event rate at Super-Kamiokande is explained in \S\ref{sec:sk}.
Our main results are presented in \S\ref{sec:results}, where we confirm that the nuclear EOS strongly affects the event rate and average event energy.
Using the backward time analysis it is possible to extract both the PNS mass and EOS.
Furthermore, we introduce a new proposal for an improved backward time analysis that incorporates information on the average event energy.
The data analysis strategy proposed in this study is summarized in \S\ref{sec:dataanl} and the detector background is described in \S\ref{sec:bkg}.
We devote \S\ref{sec:concl} to our conclusions.

\section{Nuclear Equation of State} \label{sec:eos}
The properties of nuclear matter relevant to the description of the supernova core and PNS have been studied for decades.
The incompressibility of symmetric nuclear matter, which has equal numbers of neutrons and protons, has been
deduced from measurements of the isoscalar giant monopole resonance in some nuclei \citep{2018PrPNP.101...55G}.
Heavy-ion collision data, particularly those describing the collective flow of nucleons and nuclear fragment yields,  provide constraints on the EOS at supranuclear densities
not only for symmetric nuclear matter \citep{2016NuPhA.945..112L} but also for neutron-rich matter \citep{2020arXiv201206976J}.
The density dependence of symmetric energy, which corresponds to the difference in energy per nucleon between pure neutron matter and symmetric nuclear matter,
has also been explored using observables such as nuclear masses \citep{2012PhRvL.108e2501M}, isobaric analog states \citep{2014NuPhA.922....1D},
isovector giant dipole resonances \citep{2020PhLB..81035820X}, pygmy dipole resonances \citep{2010PhRvC..81d1301C}, and the electric dipole polarizability \citep{2011PhRvL.107f2502T,2014EPJA...50...28T}.
In addition, measurements of the neutron skin thickness, defined as the difference of the neutron and proton root-mean-square radii, has been successfully applied to extract parameters characterizing the symmetry energy \citep{2000PhRvL..85.5296B,2021arXiv210103193R}.
Meanwhile, various theoretical frameworks are included in the analyses of these experiments and has helped determine the nuclear EOS to some extent \cite[see][and references therein]{2017RvMP...89a5007O,2018PrPNP.101...96R}.

In spite of this progress, developing EOS models that are applicable to simulations of supernova explosions and PNS cooling
is complicated by the need to cover a wide range of densities, temperatures, and proton fractions,
including extreme values that are out of the reach of laboratory experiments.
As a result, a variety of approaches have been proposed to extrapolate to these exotic conditions~\cite[see][for review]{2017RvMP...89a5007O}.
Among these, parameterized forms of the EOS are a convenient means of ensuring consistency with existing constraints while allowing for systematic investigations of the impact of parameter changes \citep{1985PhRvL..55..126B,2019ApJ...878...25N}.
Empirical EOS models, on the other hand, adopt effective theories to account for the average properties of terrestrial nuclei and infinite nuclear matter \citep{1991NuPhA.535..331L,1998NuPhA.637..435S,1998PThPh.100.1013S,2011ApJS..197...20S,2010NuPhA.837..210H,2013ApJ...774...17S,2019PhRvC.100e5802S,2020PTEP.2020a3D05F}.
Recently, a microscopic many-body method with realistic nuclear forces has been used to construct an EOS data table \citep{2017NuPhA.961...78T}.

In the present study we adopt four nuclear EOS models: the Shen, the LS220, the Togashi, and the T+S EOS models.
Here we briefly review their main properties.
The Shen EOS is constructed using a relativistic mean field (RMF) theory based on the Brueckner Hartree--Fock theory with a local density approximation \citep{2011ApJS..197...20S}.
Effective interactions of the RMF are determined by the properties of finite nuclei.
This EOS was also used for the neutrino event rate estimates in Paper I and in the construction of the supernova neutrino database \citep{2013ApJS..205....2N}.
On the other hand, the LS220 EOS is one of the Lattimer--Swesty EOS sets provided by \citet{1991NuPhA.535..331L} and has an incompressibility of 220 MeV.
These sets are constructed with an energy function of Skyrme-type nuclear interactions and a model of nuclei in the compressible liquid drop model.
The parameters of the energy function are determined by the bulk properties of nuclear matter saturation.
The LS220 EOS has been used in several supernova simulations~\citep[see, e.g.,][for the EOS impact on supernova dynamics]{suwa13a,fisc14,hara20}.
The Togashi EOS is based on the cluster variational method starting from a realistic nucleon--nucleon potential and local density approximation \citep{2017NuPhA.961...78T}.
Nuclear interactions are determined by data from scattering experiments and supplemented by the empirical properties of nuclear matter at the saturation density obtained from other experiments.
The T+S EOS adopts a hybrid approach using the Togashi EOS at high densities and the Shen EOS at low densities to examine the effect of the nonuniform phase of hot and dense matter.
Indeed, heavy nuclei residing near the surface of the PNS affect neutrino transport and the resulting neutrino emission \citep{2018PhRvC..97c5804N}.
Note that a single spherical heavy nucleus is assumed to be located at the center of the Wigner-Seitz cell of a body-centered cubic lattice for both the Shen EOS and Togashi EOS models,
though the mass number of the nuclei is larger in the Togashi EOS.

\begin{figure}[t!]
\epsscale{0.5}
\plotone{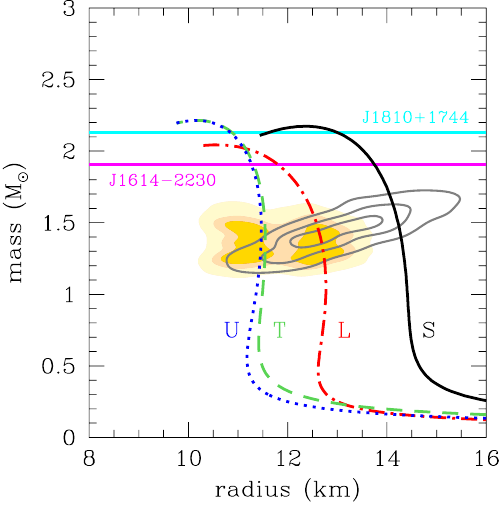}
\caption{Mass--radius relations of cold neutron stars for the EOS models adopted in this study. Solid (black), dot-dashed (red), dashed (green), and dotted (blue) curves are for the Shen EOS (S), LS220 EOS (L), Togashi EOS (T), and T+S EOS (U), respectively. The T+S EOS is constructed by connecting the Togashi EOS at high densities and the Shen EOS at low densities including the nonuniform phase. Filled and open contours correspond to constraints from the gravitational wave and X-ray observations, respectively. Horizontal lines represent mass measurements of heavy neutron stars.
\label{fig:mr}}
\end{figure}

In Figure \ref{fig:mr} we plot the gravitational mass and radius of cold neutron stars for the EOS models adopted in this study.
Note that the Togashi EOS and T+S EOS have similar mass--radius relations because the structure of neutron stars is mainly determined by the high-density part of the EOS,
which is the same for these models.
In contrast, the results with the Shen, LS220, and Togashi models have different radii due to their differing theoretical approaches and experimental inputs.
Since the Shen EOS is the stiffest of these, it shows the largest radius, and the softer Togashi EOS results in a smaller radius.
Observational constraints from the LIGO--Virgo detection of GW170817 \citep{2018PhRvL.121p1101A}, the mass measurements of heavy neutron stars PSR J1614-2230 \citep{2018ApJS..235...37A} and PSR J1810+1744 \citep{2021arXiv210109822R}\footnote{See \citet{2013Sci...340..448A} and \citet{2020NatAs...4...72C} for other heavy neutron stars.}, and the {\it NICER}  observation of isolated millisecond pulsar PSR J0030+0451 \citep{2019ApJ...887L..24M} are also shown in the figure\footnote{Recently {\it NICER} and {\it XMM-NEWTON}  report that the neutron star PSR J0740+6620 has a mass of $2.08^{+0.07}_{-0.07}M_\odot$ and a radius of $13.7^{+2.6}_{-1.5}$~km \citep{2021arXiv210506979M}. This result implies a larger radius than the previous work and further investigation is important.}.
With the exception of the Shen EOS, which
is not compatible with the gravitational wave data\footnote{The authors of the Shen EOS have recently presented a new EOS table using the same theoretical approach but with updated symmetry energy terms so to better match the experimental constraints \citep{2020ApJ...891..148S}. Its influence on the neutrino signals is reported in \citet{2019ApJ...887..110S}.},
the EOS models adopted in this study are roughly consistent with these observations.

\section{Setup for Event Rate Estimations} \label{sec:setup}
\subsection{Modeling Neutrino Emission from PNS Cooling} \label{sec:pns}
In order to model the supernova neutrino emission in the late phase, we perform PNS cooling simulations. 
As in Paper I we utilize the general relativistic quasi-static evolutionary code that solves simultaneously the PNS structure (Oppenheimer-Volkoff equation) and neutrino transfer using a Henyey-type method assuming spherical symmetry \citep{1994pan..conf..763S}.
Note that spherically symmetric hydrodynamics codes have been employed for the PNS cooling simulations in other recent studies \citep{2021PhRvD.103b3016L,2020arXiv201016254M}.
For neutrino transfer a multigroup flux-limited scheme and the flux limiter in \citet{1987ApJ...318..288M} are adopted in our model. 
We follow neutrino interactions with matter taking into account the energy dependence of $\nu_e$, $\bar\nu_e$, and $\nu_x$, where $\nu_\mu$, $\bar\nu_\mu$, $\nu_\tau$, and $\bar\nu_\tau$ are treated collectively as $\nu_x$ \citep{2019ApJ...878...25N}. 
Consequently, we obtain the thermal evolution and neutrino emission of the PNS. For simplicity the effects of additional mass accretion \citep{1988ApJ...334..891B} and convection \citep{2012PhRvL.108f1103R} are not considered.
In particular, convection smooths the entropy gradient and accelerates cooling in the early phase. Note however, that this does not change our results for the late phase much. According to Paper I, although the initial entropy affects the time of the last expected event, the backward cumulative event number measured from the time of the last expected event is insensitive to the initial entropy. Since convection changes the entropy profile, the impact of convection would be analogous to that of the initial entropy profile. Convection is also ignored in \citet{2021PhRvD.103b3016L} and \citet{2020arXiv201016254M}.
Note that while the neutrino luminosity in \citet{2018PhRvC..97c5804N} had numerical fluctuations for the Togashi EOS case, the neutrino light curves obtained in this paper become smooth after modifications to the code.\footnote{In the PNS cooling code the discrete EOS table data are interpolated. The previous interpolation over the fraction of atomic nuclei caused a large error in the region with a large mass number, which we have now improved to be consistent with baryon number conservation.}

Due to neutrino emission the entropy and lepton number in the PNS evolve.
We set the entropy and electron fraction profiles and construct hydrostatic configurations with an almost steady flow of neutrinos to use as the initial conditions of the PNS cooling simulations.
The initial entropy and electron fraction profiles are taken from  snapshots of the hydrodynamical simulation as functions of the baryon mass coordinate.
For this purpose we perform core-collapse simulations of the progenitors with a zero-age main sequence mass of $M_{\rm ZAMS}=15 M_\odot$ and $40 M_\odot$ \citep{1995ApJS..101..181W} using the general relativistic neutrino radiation hydrodynamics code \citep{2005ApJ...629..922S}. We follow the core collapse adopting the Shen EOS, LS220 EOS, and Togashi EOS individually for the initial conditions of the PNS cooling simulations. The initial conditions of the PNS cooling with the T+S EOS models are taken from the core collapse with the Togashi EOS.

\begin{figure*}[t]
\gridline{\fig{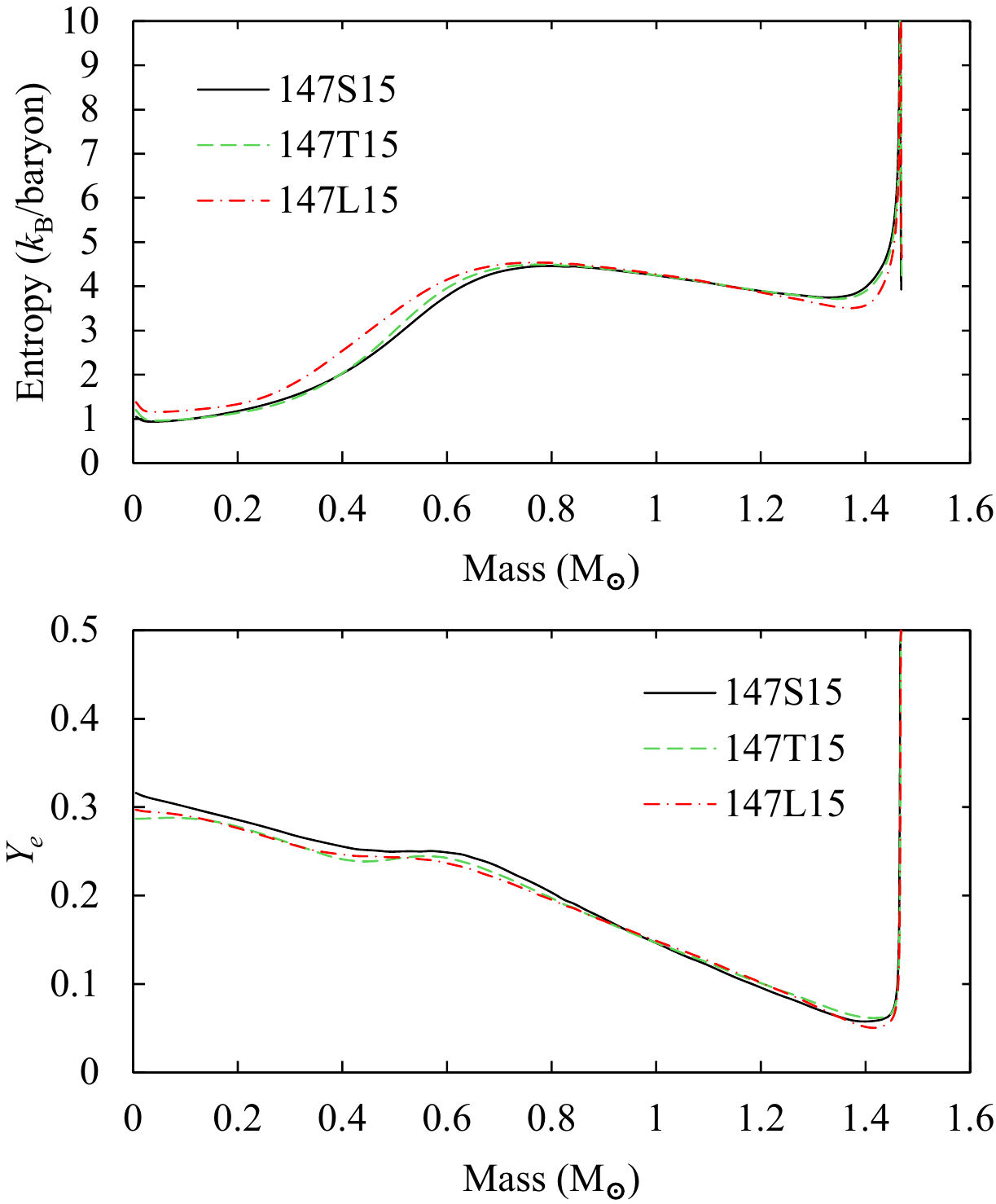}{0.33\textwidth}{(a) $M_b=1.47 M_\odot$}
          \fig{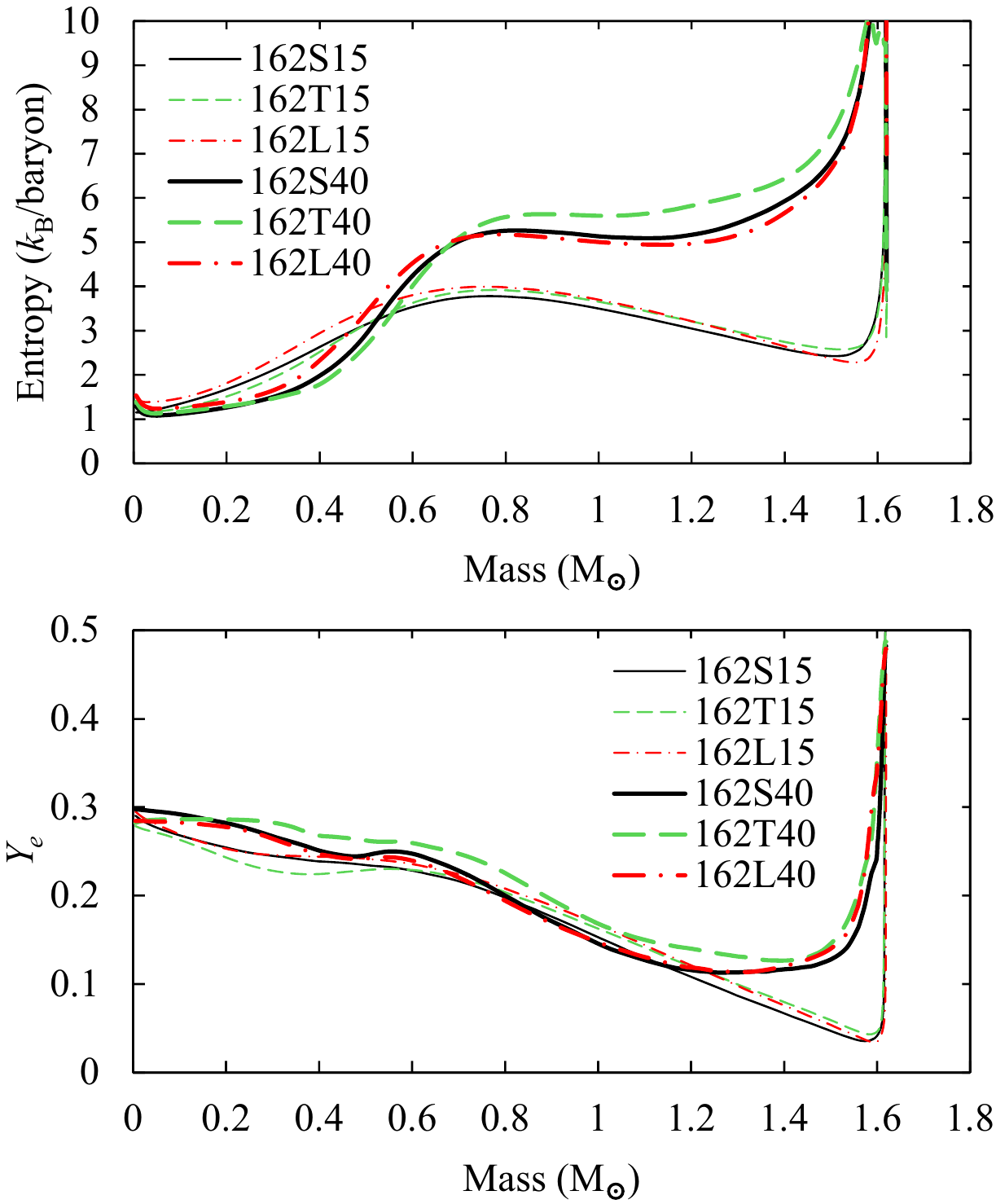}{0.33\textwidth}{(b) $M_b=1.62 M_\odot$}
          \fig{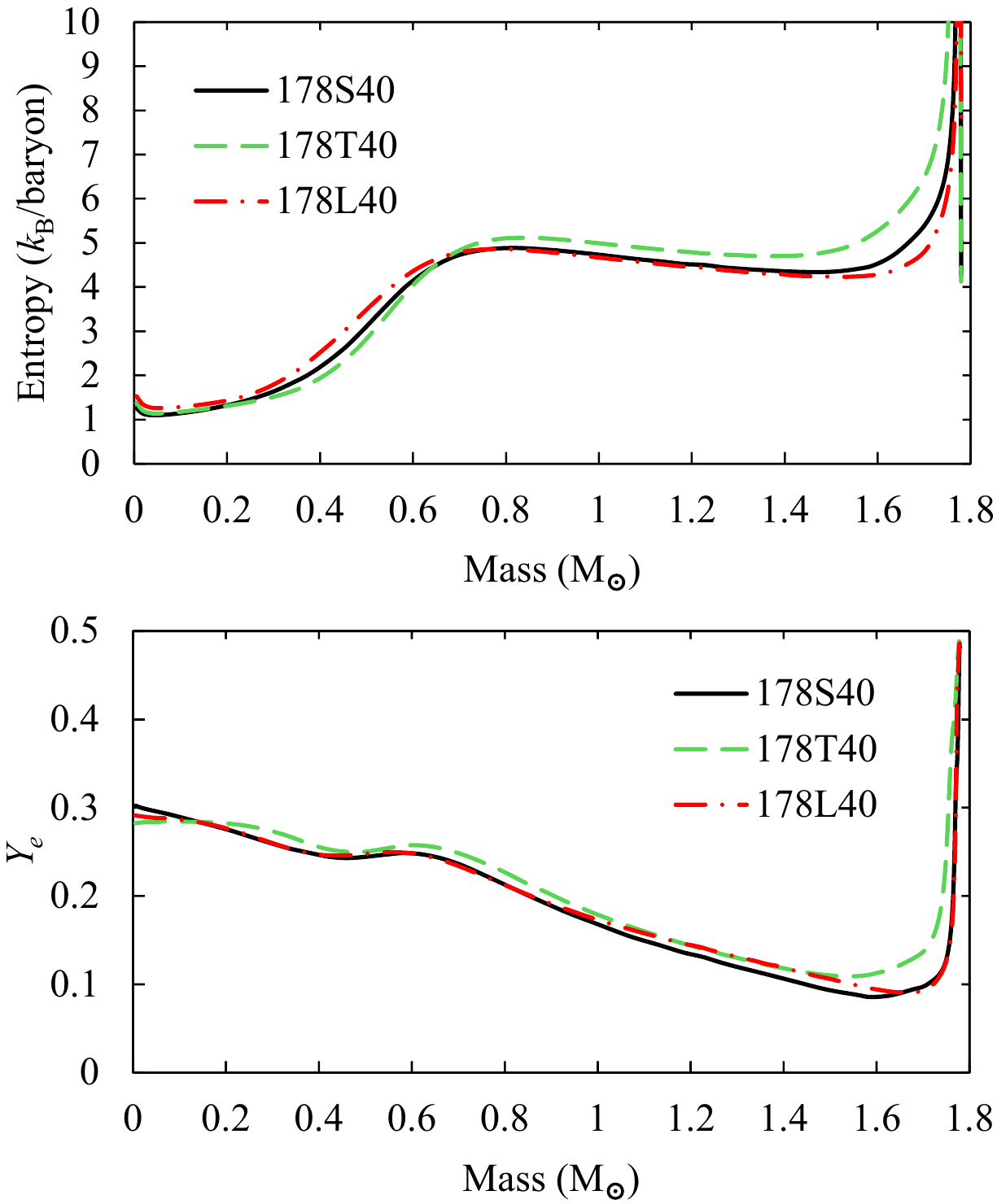}{0.33\textwidth}{(c) $M_b=1.78 M_\odot$}
        }
\caption{Initial entropy profile (upper) and electron fraction profile (lower) as functions of the baryon mass coordinate.
Thin and thick lines show the results of core-collapse simulations for models with $M_{\rm ZAMS}=15 M_\odot$ and $40 M_\odot$, respectively. Solid (black), dashed (green), and dot-dashed (red) lines correspond to models with the Shen EOS, the Togashi EOS, and the LS220 EOS, respectively.
\label{fig:inicon}}
\end{figure*}

The models of PNS cooling investigated in this paper are listed in Table~\ref{tab:enumber}. 
For instance, in the core-collapse simulation of the $15 M_\odot$ progenitor with the Shen EOS, the shock wave is located at the baryon mass coordinate $1.40 M_\odot$ 110 ms after the bounce. 
Next we extract the central part of the stellar core up to just ahead of the shock wave and start the PNS cooling simulation of the 140S15 model.
This model has a baryon mass of $M_b=1.40 M_\odot$ and the Shen EOS from $t_{\rm init}=110$~ms after the bounce. 
In this paper the models are denoted as xxxYzz where xxx and zz correspond  to $M_b$ and $M_{\rm ZAMS}$, respectively, and Y = S, L, T, U denotes the Shen EOS, LS220 EOS, Togashi EOS, and T+S EOS, respectively.
Note that the 147S15 model is identical to the 147S model in Paper I. 
In Figure~\ref{fig:inicon} the initial entropy and electron fraction profiles are shown for some models. 
Comparing the eight models with $M_b=1.62 M_\odot$ we see that the difference in the progenitor model is larger than that in the EOS especially for the entropy profile. Incidentally, the profiles of models not shown in Figure~\ref{fig:inicon} are similar to those with the same EOS and progenitor models.
For instance, since the models 140S15--162S15 are based on snapshots of the same core-collapse simulation at different times, the profiles of the model 154S15 is between that of the models 147S15 and 162S15.
\begin{deluxetable*}{lclcrcccrrrrr}
\tablecaption{List of PNS cooling models. \label{tab:enumber}}
\tablewidth{0pt}
\tablehead{
\colhead{Model} & \colhead{$M_{\rm ZAMS}$} & \colhead{EOS} & \colhead{$M_b$} & \colhead{$t_{\rm init}$} & \colhead{$M_{\rm NS,g}$} & \colhead{$R_{\rm NS}$} & \colhead{$\Delta M_{\rm g}$} & \colhead{$N_{\rm tot}$} & \colhead{$N(>\!1\,{\rm s})$} & \colhead{$N(>\!10\,{\rm s})$} & \colhead{$N(>\!20\,{\rm s})$} & \colhead{$t_{\rm last}$} \\
\colhead{} & \colhead{($M_\odot$)} & \colhead{} & \colhead{($M_\odot$)} & \colhead{(s)} & \colhead{($M_\odot$)} & \colhead{(km)} & \colhead{($M_\odot$)} & \colhead{} & \colhead{} & \colhead{} & \colhead{} & \colhead{(s)}
}
\startdata
140S15 & 15 & Shen & 1.40 & 0.110 & 1.289 & 14.33 & 0.094 & 2,560 & 1,644 & 413 & 110 & 42.0 \\
147S15 & 15 & Shen & 1.47 & 0.300 & 1.348 & 14.31 & 0.090 & 2,241 & 1,805 & 499 & 148 & 45.2 \\
154S15 & 15 & Shen & 1.54 & 0.602 & 1.407 & 14.28 & 0.089 & 2,176 & 1,966 & 609 & 202 & 49.0 \\
162S15 & 15 & Shen & 1.62 & 1.012 & 1.473 & 14.24 & 0.088 & 2,117 & 2,117 & 719 & 264 & 53.1 \\
162S40 & 40 & Shen & 1.62 & 0.092 & 1.473 & 14.24 & 0.129 & 3,555 & 2,440 & 796 & 294 & 54.2 \\
170S40 & 40 & Shen & 1.70 & 0.145 & 1.539 & 14.20 & 0.136 & 3,673 & 2,714 & 955 & 382 & 59.0 \\
178S40 & 40 & Shen & 1.78 & 0.206 & 1.604 & 14.14 & 0.141 & 3,828 & 2,974 & 1,124 & 482 & 63.9 \\
186S40 & 40 & Shen & 1.86 & 0.274 & 1.668 & 14.08 & 0.147 & 4,216 & 3,221 & 1,303 & 595 & 69.3 \\ \hline
140L15 & 15 & LS220 & 1.40 & 0.133 & 1.277 & 12.73 & 0.106 & 2,946 & 2,144 & 803 & 362 & 69.4 \\
147L15 & 15 & LS220 & 1.47 & 0.325 & 1.335 & 12.70 & 0.105 & 2,769 & 2,369 & 955 & 458 & 75.7 \\
154L15 & 15 & LS220 & 1.54 & 0.642 & 1.392 & 12.66 & 0.104 & 2,695 & 2,528 & 1,106 & 563 & 82.4 \\
162L15 & 15 & LS220 & 1.62 & 1.061 & 1.457 & 12.62 & 0.101 & 2,575 & 2,575 & 1,233 & 664 & 89.6 \\
162L40 & 40 & LS220 & 1.62 & 0.110 & 1.457 & 12.62 & 0.145 & 4,385 & 3,112 & 1,401 & 751 & 92.0 \\
170L40 & 40 & LS220 & 1.70 & 0.166 & 1.521 & 12.56 & 0.153 & 4,598 & 3,439 & 1,642 & 924 & 100.4 \\
178L40 & 40 & LS220 & 1.78 & 0.230 & 1.584 & 12.49 & 0.159 & 4,411 & 3,742 & 1,893 & 1,116 & 110.6 \\
186L40 & 40 & LS220 & 1.86 & 0.299 & 1.647 & 12.40 & 0.166 & 4,588 & 4,007 & 2,147 & 1,322 & 122.0 \\ \hline
140T15 & 15 & Togashi & 1.40 & 0.105 & 1.266 & 11.54 & 0.120 & 3,531 & 2,578 & 1,192 & 685 & 126.5 \\
147T15 & 15 & Togashi & 1.47 & 0.300 & 1.323 & 11.55 & 0.118 & 3,289 & 2,843 & 1,382 & 823 & 132.5 \\
154T15 & 15 & Togashi & 1.54 & 0.605 & 1.379 & 11.55 & 0.121 & 3,338 & 3,115 & 1,602 & 989 & 139.0 \\
162T15 & 15 & Togashi & 1.62 & 0.974 & 1.443 & 11.55 & 0.121 & 3,605 & 3,291 & 1,802 & 1,153 & 146.0 \\
162T40 & 40 & Togashi & 1.62 & 0.061 & 1.443 & 11.55 & 0.166 & 5,075 & 3,705 & 1,940 & 1,235 & 148.0 \\
170T40 & 40 & Togashi & 1.70 & 0.103 & 1.505 & 11.54 & 0.178 & 5,382 & 4,124 & 2,248 & 1,474 & 155.8 \\
178T40 & 40 & Togashi & 1.78 & 0.146 & 1.567 & 11.53 & 0.187 & 5,560 & 4,533 & 2,568 & 1,731 & 164.4 \\
186T40 & 40 & Togashi & 1.86 & 0.214 & 1.628 & 11.51 & 0.196 & 5,907 & 4,944 & 2,908 & 2,012 & 173.7 \\ \hline
140U15 & 15 & T+S & 1.40 & 0.105 & 1.266 & 11.45 & 0.119 & 3,298 & 2,357 & 990 & 496 & 69.3 \\
147U15 & 15 & T+S & 1.47 & 0.300 & 1.323 & 11.46 & 0.116 & 3,004 & 2,569 & 1,148 & 604 & 74.7 \\
154U15 & 15 & T+S & 1.54 & 0.605 & 1.379 & 11.47 & 0.118 & 3,013 & 2,794 & 1,329 & 736 & 80.7 \\
162U15 & 15 & T+S & 1.62 & 0.974 & 1.442 & 11.47 & 0.120 & 3,256 & 2,955 & 1,502 & 873 & 87.1 \\
162U40 & 40 & T+S & 1.62 & 0.061 & 1.442 & 11.47 & 0.165 & 4,732 & 3,379 & 1,637 & 947 & 89.0 \\
170U40 & 40 & T+S & 1.70 & 0.103 & 1.505 & 11.47 & 0.176 & 4,984 & 3,743 & 1,899 & 1,144 & 96.7 \\
178U40 & 40 & T+S & 1.78 & 0.146 & 1.567 & 11.46 & 0.184 & 5,094 & 4,084 & 2,166 & 1,352 & 104.6 \\
186U40 & 40 & T+S & 1.86 & 0.214 & 1.628 & 11.44 & 0.193 & 5,368 & 4,421 & 2,447 & 1,579 & 113.3 \\
\enddata
\tablecomments{Models considered in this paper, where $M_{\rm ZAMS}$ is the zero-age main sequence mass of the progenitor model and $M_b$ is the baryon mass of the PNS, which is invariant throughout the cooling process. 
Here $t_{\rm init}$ is the time when the initial condition of the PNS cooling simulation is extracted from the core-collapse simulation as measured from the core bounce, $t_{\rm pb}$. The parameters $M_{\rm NS,g}$ and $R_{\rm NS}$ are the gravitational mass and radius of the neutron star born in the supernova explosion, respectively. 
Similarly, $\Delta M_{\rm g}$ is the difference between the initial gravitational mass of the PNS and that of the cold neutron star with the same baryon mass. Here $N_{\rm tot}$ is the number of events from the PNS cooling only and does not include the preceding stage.
This total is broken down into $N(>\!1\,{\rm s})$, $N(>\!10\,{\rm s})$, and $N(>\!20\,{\rm s})$, which represent the numbers of events expected after $t_{\rm pb}=1$~s, 10~s, and 20~s, respectively, as defined in Eq.~(\ref{eq:revcumev}). The time of the last expected event is $t_{\rm last}$. 
The distance to the supernova is assumed to be 10~kpc for the evaluations of the event numbers and $t_{\rm last}$.}
\end{deluxetable*}

\begin{figure}[t!]
\epsscale{0.5}
\plotone{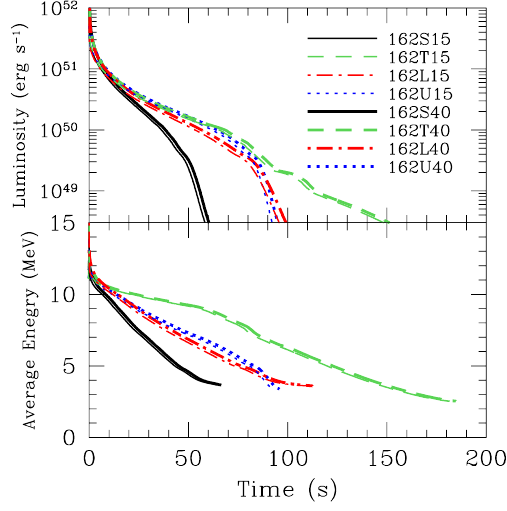}
\caption{Luminosity (upper) and average energy (lower) of $\bar\nu_e$ as a function of time after the bounce for PNS models with a baryon mass of $M_b=1.62 M_\odot$. The lines have the same meaning as in Figure~\ref{fig:inicon} except for the dotted lines, which  correspond to models with the T+S EOS.
\label{fig:nuLC162}}
\end{figure}

\begin{figure*}[t]
\gridline{\fig{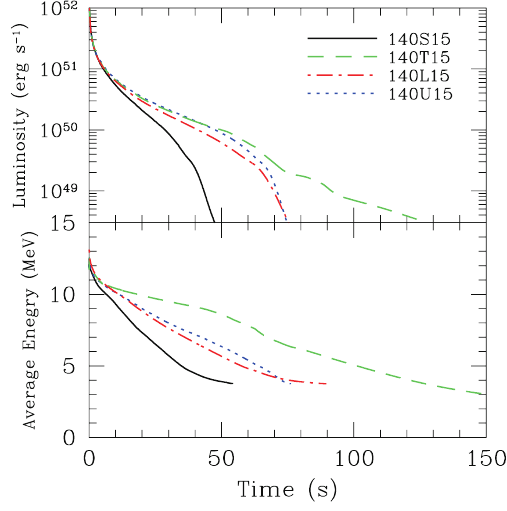}{0.33\textwidth}{(a) $M_b=1.40 M_\odot$}
          \fig{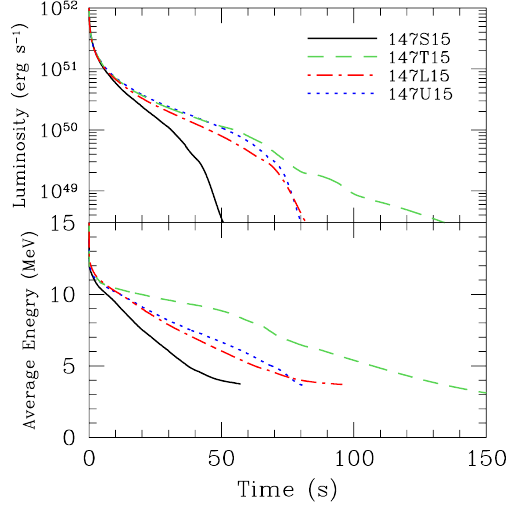}{0.33\textwidth}{(b) $M_b=1.47 M_\odot$}
          \fig{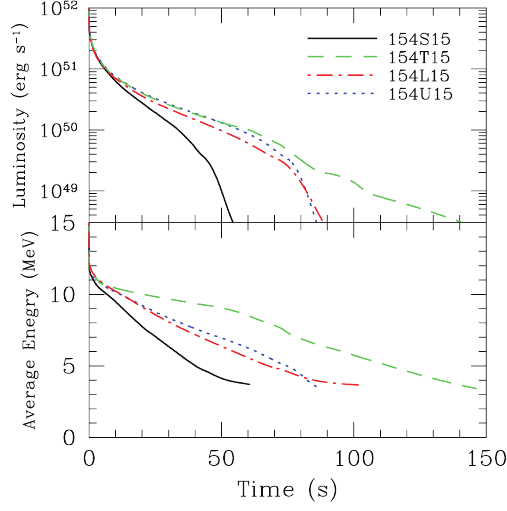}{0.33\textwidth}{(c) $M_b=1.54 M_\odot$}
          }
\gridline{\fig{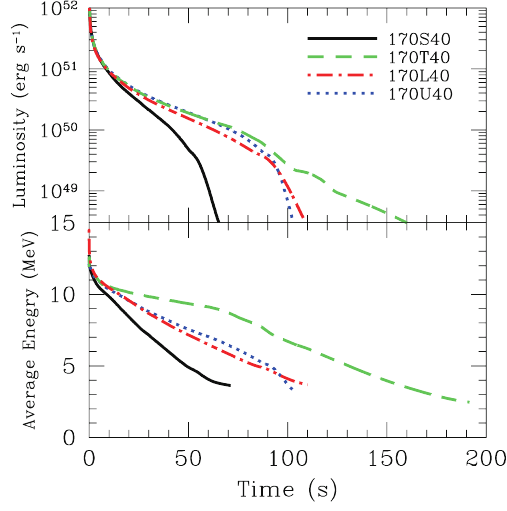}{0.33\textwidth}{(d) $M_b=1.70 M_\odot$}
          \fig{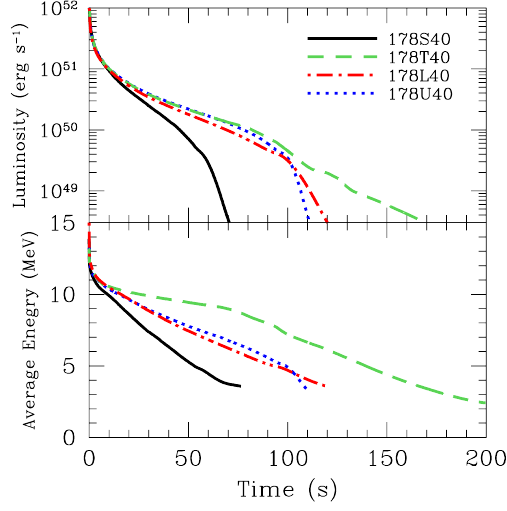}{0.33\textwidth}{(e) $M_b=1.78 M_\odot$}
          \fig{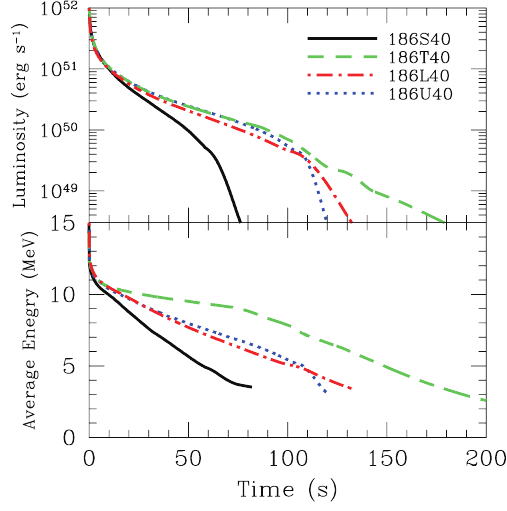}{0.33\textwidth}{(f) $M_b=1.86 M_\odot$}
          }
\caption{Same as Figure~\ref{fig:nuLC162} but for PNS models with a baryon mass $M_b$ of (a) $1.40 M_\odot$, (b) $1.47 M_\odot$, (c) $1.54 M_\odot$, (d) $1.70 M_\odot$, (e) $1.78 M_\odot$, and (f) $1.86 M_\odot$. 
\label{fig:nuLC}}
\end{figure*}

In Figures~\ref{fig:nuLC162} and \ref{fig:nuLC}, we show the $\bar\nu_e$ luminosity and average energy of the PNS cooling models considered in this paper. 
Provided that the EOS is the same, the models with a higher PNS mass show longer neutrino emission. 
As reported in Paper I, the difference in the initial entropy has only a minor impact on the neutrino signal (Figure~\ref{fig:nuLC162}). 
In contrast, the EOS affects the neutrino emission from PNS cooling.
Until the $\bar\nu_e$ luminosity drops to $\sim$10$^{50}$~erg~s$^{-1}$ the model with the T+S EOS has a similar neutrino light curve to that of the Togashi EOS for each PNS mass. 
In this period, the decay timescale of the neutrino light curve\footnote{In \citet{2019ApJ...878...25N}, the cooling timescale is defined by the maximum $e$-folding time of  the $\bar\nu_e$ luminosity.}, $\tau$, is determined by the EOS at supranuclear densities for which the Togashi EOS and T+S EOS models are the same. Furthermore, as derived from the Kelvin--Helmholtz timescale with general relativity, it is longer for models with a smaller neutron star radius, $R_{\rm NS}$, as \citep{2020ApJ...891..156N}
\begin{equation}
\tau \propto \frac{M_{\rm NS,g}^2}{R_{\rm NS}^3 (1-0.5\beta) \sqrt{1-2\beta}} ,
\label{eq:taucool}
\end{equation}
with
\begin{equation}
\beta = \frac{GM_{\rm NS,g}}{c^2 R_{\rm NS}} ,
\label{eq:compactness}
\end{equation}
where $G$ and $c$ are the gravitational constant and the velocity of light, respectively. 
The EOS dependence of the models in this study is consistent with this trend (Table~\ref{tab:enumber}); the models with the Shen EOS, whose $R_{\rm NS}$ is larger, have a shorter timescale.
Note that the gravitational mass of the cold neutron star, $M_{\rm NS,g}$, is determined by the EOS once $M_b$ is given (Table~\ref{tab:enumber}).

After the Kelvin--Helmholtz cooling phase, the $\bar\nu_e$ luminosity drops rapidly because the PNS has become transparent to neutrinos and the previously trapped neutrinos have diffused out.
A similar falloff in the neutrino luminosity is also seen in the model of \citet{2021PhRvD.103b3016L}.
Meanwhile, the difference in the neutrino light curve gets larger between the Togashi EOS model and T+S EOS model. 
Furthermore, the Togashi EOS model has a higher average energy than the T+S EOS model at earlier times. 
These differences are attributed to the EOS at subnuclear densities.
Indeed, since heavy nuclei near the PNS surface have a large mass number for the Togashi EOS models, they have a large coherent scattering cross section with neutrinos. 
Accordingly, the efficiency of the electron scattering reaction, which associates the energy exchange between matter and neutrinos, is enhanced and thermalization is achieved at lower densities \citep{2018PhRvC..97c5804N}.
This does not change the neutrino light curve much but it does affect the thermal structure of PNS near the surface. 
Therefore, in comparison with the T+S EOS models, the Togashi EOS models have a higher temperature near the PNS surface and a higher average energy of neutrinos emitted from the PNS.

\subsection{Neutrino Detection at Super-Kamiokande} \label{sec:sk}
We estimate the event rates of supernova neutrinos at Super-Kamiokande as was done in Paper I. 
When uncertainties in the nuclear EOS are taken into account the numerical simulations of PNS cooling show a wide variety of neutrino light curves.
We aim to update our analysis method proposed in Paper I so as to cover the various types of neutrino signals. 
For this purpose, we focus on the inverse beta decay (IBD) interactions \citep{2003PhLB..564...42S}:
\begin{equation}
\bar \nu_e + p \to e^+ + n ,
\label{eq:ibd}
\end{equation}
which is the most dominant channel in water Cherenkov detectors such as Super-Kamiokande.
In 2020 gadolinium (Gd) was loaded in the Super-Kamiokande detector's water \citep{2004PhRvL..93q1101B,2017JPhCS.888a2041S,Marti-Magro:20210w,2021arXiv210900360A} starting the SK-Gd period of operations.
This upgrade allows for the efficient detection of neutrons, further enabling IBD events to be separated from both electron-scattering \citep{1995PhRvD..51.6146B} and $^{16}$O charged-current events \citep{2018arXiv180908398N}.
In this paper we estimate IBD event rates assuming perfect neutron tagging. We leave investigations of the impact of other interaction channels and the neutron tagging efficiency to future study \citep[see][]{2014PhRvD..89f3007L}.
A detailed description of Super-Kamiokande is provided in \citet{2003NIMPA.501..418F}.

The spectrum of IBD events per unit time is estimated as
\begin{equation}
\frac{d^2N(E_{e^+},t)}{dE_{e^+}dt}=N_T\int^1_{-1} d\cos\theta \frac{d\sigma(E_\nu,\cos\theta)}{d\cos\theta}\frac{d\phi_{\bar\nu_e}(E_\nu,t)}{dE_\nu} \left( \frac{\partial E_\nu}{\partial E_{e^+}}\right)_{\cos\theta},
\label{eq:evratesp}
\end{equation}
where $N_T$ is the number of proton targets in Super-Kamiokande. 
In this paper we assume that the entire 32.5~kton volume of the inner detector\footnote{While the volume is listed as 32 kton in \citet{2003NIMPA.501..418F}, the precise volume is 32.481~kton. The two are close enough for our purposes, so we adopt 32.5~kton for convenience as was done in Paper I.} is used for $N_T$. 
For the $\bar\nu_e$ flux spectrum, $d\phi_{\bar\nu_e}/dE_\nu$,
we assume that the neutrinos described in \S~\ref{sec:pns} are emitted from a supernova exploding at a distance of 10~kpc. 
We utilize the IBD cross section $d\sigma/d\cos\theta$ calculated by \citet{2003PhLB..564...42S}.\footnote{In Paper I, we adopted the cross section in \citet{1999PhRvD..60e3003V}, which is similar to \citet{2003PhLB..564...42S} below 40~MeV. See also Table 2 of \citet{2003PhLB..564...42S}.}
As seen in Eq.~(21) of the reference, the incoming neutrino energy $E_\nu$ can be expressed as a function of the energy of recoil positrons $E_{e^+}$ and the scattering angle $\theta$. The positron rest mass is included in $E_{e^+}$.
Furthermore, the detector response and energy resolution of Super-Kamiokande are taken into account in our calculation \citep[][]{2016arXiv160607538S}, though their impact on this analysis is not significant. 
Substituting these factors into Eq.~(\ref{eq:evratesp}), we obtain the event rate per unit time:
\begin{equation}
\dot{N}(t)=\int^\infty_{E_\mathrm{th}} dE_{e^+}\frac{d^2N(E_{e^+},t)}{dE_{e^+}dt}.
\label{eq:evrate}
\end{equation}
Here $E_\mathrm{th}$ is the threshold energy of positron detection at Super-Kamiokande, which we set to $E_\mathrm{th}=5$~MeV in this study. 
Incidentally, in \citet{2021PhRvD.103b3016L}, the threshold is set to be 4.0~MeV total positron energy (3.5~MeV kinetic energy) and the detector mass is assumed to be the fiducial volume (FV), 22.5~kton, as is the case for Super-Kamiokande's solar-neutrino studies. Furthermore, the reverse cumulative event number, which is essential for our backward time analysis, is given by
\begin{equation}
N(>\!t)=\int^\infty_t dt^\prime \dot{N}(t^\prime).
\label{eq:revcumev}
\end{equation}
To investigate the impact of the nuclear EOS, we also calculate the average energy of recoil positrons with
\begin{equation}
\bar E_{e^+}(t)= \frac{1}{\dot{N}(t)}\int^\infty_{E_\mathrm{th}} dE_{e^+} \, E_{e^+}\frac{d^2N(E_{e^+},t)}{dE_{e^+}dt},
\label{eq:aveene}
\end{equation}
which also reflects the average energy of the neutrinos.

\section{Results} \label{sec:results}
\subsection{Expected Event Rate} \label{sec:eventrate}
\begin{figure}[t!]
\epsscale{0.5}
\plotone{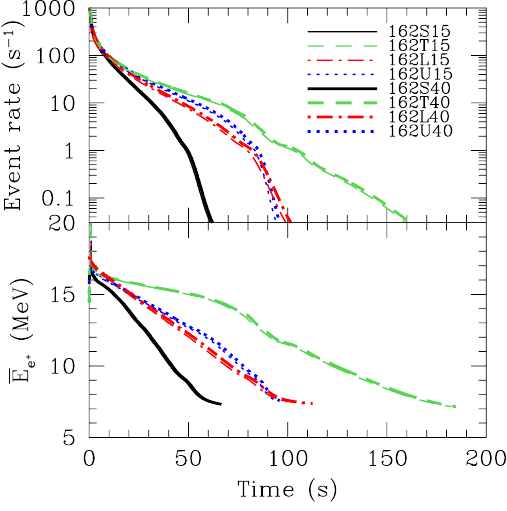}
\caption{Event rate (upper) and average energy (lower) of positrons from IBD reactions as a function of time after the bounce for PNS models with a baryon mass of $M_b=1.62 M_\odot$. The meaning of the lines is the same as in Figure~\ref{fig:nuLC162}.
\label{fig:evev162}}
\end{figure}

\begin{figure*}[t]
\gridline{\fig{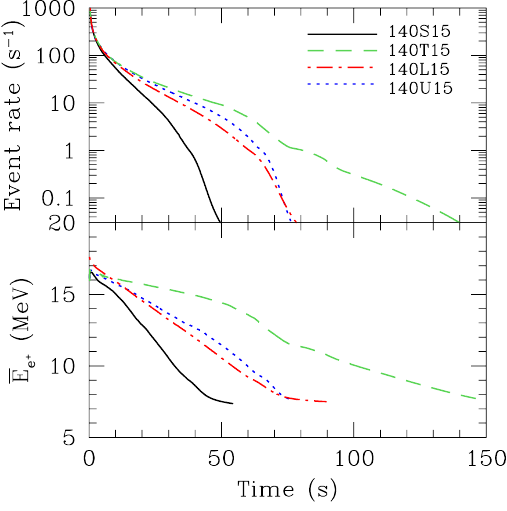}{0.33\textwidth}{(a) $M_b=1.40 M_\odot$}
          \fig{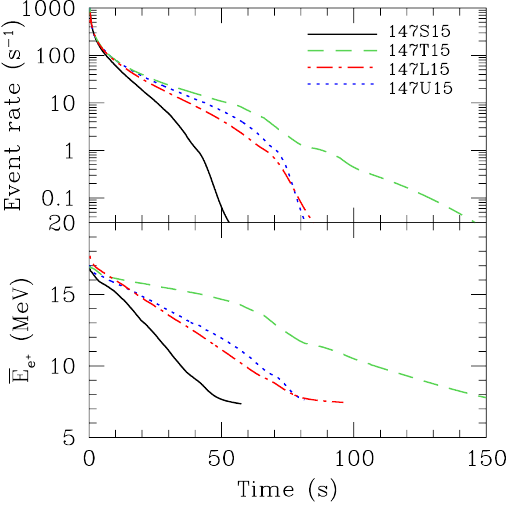}{0.33\textwidth}{(b) $M_b=1.47 M_\odot$}
          \fig{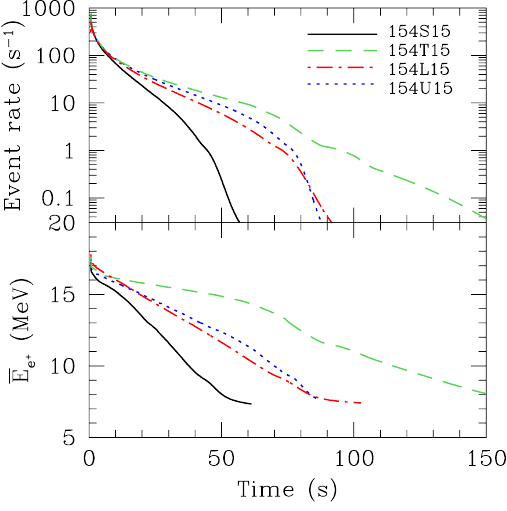}{0.33\textwidth}{(c) $M_b=1.54 M_\odot$}
          }
\gridline{\fig{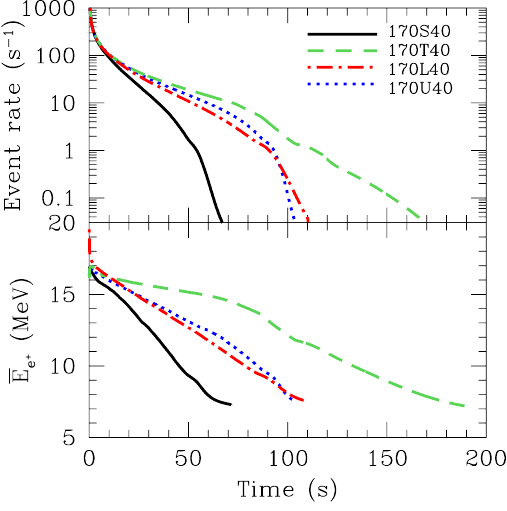}{0.33\textwidth}{(d) $M_b=1.70 M_\odot$}
          \fig{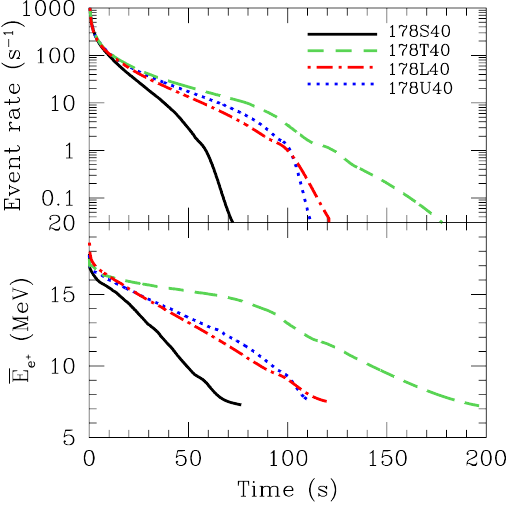}{0.33\textwidth}{(e) $M_b=1.78 M_\odot$}
          \fig{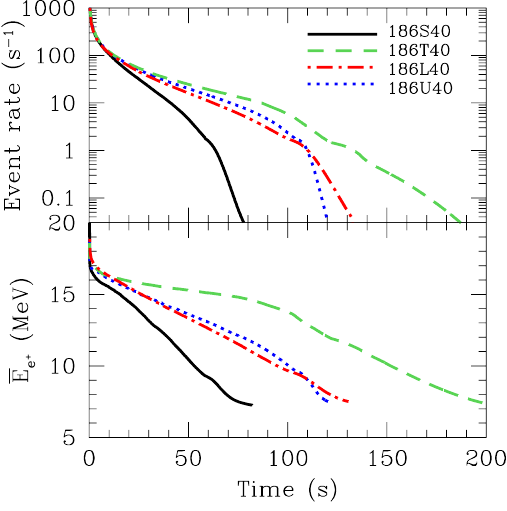}{0.33\textwidth}{(f) $M_b=1.86 M_\odot$}
          }
\caption{Same as Figure~\ref{fig:evev162} but for PNS models with a baryon mass $M_b$ of (a) $1.40 M_\odot$, (b) $1.47 M_\odot$, (c) $1.54 M_\odot$, (d) $1.70 M_\odot$, (e) $1.78 M_\odot$, and (f) $1.86 M_\odot$. 
\label{fig:evev}}
\end{figure*}

In Figures \ref{fig:evev162} and \ref{fig:evev} we show the expected event rate, $\dot{N}(t_{\rm pb})$, and average energy of recoil positrons, $\bar E_{e^+}(t_{\rm pb})$, for IBD events.
Here, $t_{\rm pb}$ is the time measured from the bounce. 
The time evolution of $\bar E_{e^+}$ is similar to that of the $\bar \nu_e$ average energy for each model except that the recoil positrons have a higher energy than the original neutrinos (Figures \ref{fig:nuLC162} and \ref{fig:nuLC}). 
This is because the IBD cross section is higher for neutrinos with higher energy and only events with energy higher than the threshold are included. 
On the other hand, the event rate is determined by both the $\bar \nu_e$ luminosity and average energy. 
For instance, the $\bar \nu_e$ luminosity in the Kelvin--Helmholtz phase, which is characterized by Eq.~(\ref{eq:taucool}), is similar for models with the Togashi EOS and T+S EOS provided that the PNS mass is the same. 
In contrast, the Togashi EOS models have a higher event rate than those of the T+S EOS models due to the higher $\bar \nu_e$ average energy. Therefore, the event rate in the Kelvin--Helmholtz phase depends not only on $R_{\rm NS}$ but also on the composition near the PNS surface.

\begin{figure}[t!]
\epsscale{0.5}
\plotone{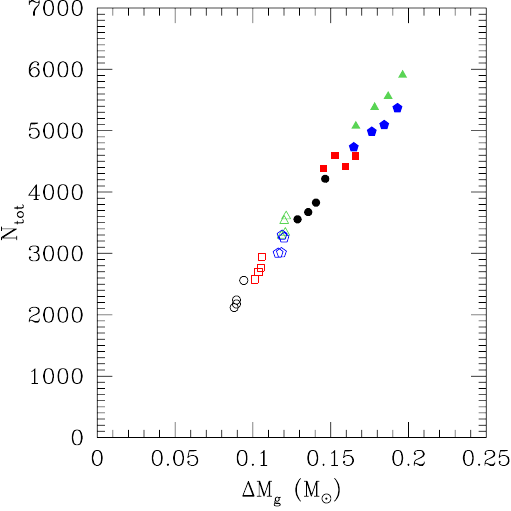}
\caption{Relation between $N_{\rm tot}$ and $\Delta M_{\rm g}$, where $N_{\rm tot}$ is the total number of IBD events and $\Delta M_{\rm g}$ is the difference between the initial gravitational mass of the PNS and that of the cold neutron star with the same baryon mass. Circles, squares, pentagons, and triangles correspond to  models with the Shen EOS, the LS220 EOS, the T+S EOS, and the Togashi EOS, respectively. Open and filled plots are for models with $M_{\rm ZAMS}=15$ and $40M_\odot$, respectively.
\label{fig:ntot}}
\end{figure}

For all models the total event number, $N_{\rm tot}$, is listed in Table~\ref{tab:enumber} with $N(>\!1\,{\rm s})$, $N(>\!10\,{\rm s})$, and $N(>\!20\,{\rm s})$, whose definition is given in Eq.~(\ref{eq:revcumev}).
Note that contributions from the preceding stages, such as a core bounce and accretion phases, are not taken into account in $N_{\rm tot}$ (See Table 1 in Paper I).
The models with the Togashi EOS, which have the highest event rates and longest cooling timescales, have the largest event numbers.
They are followed by models with the T+S EOS, LS220 EOS and Shen EOS, in that order. 
The event number difference gets larger at later times because the event rate drops rapidly after the Kelvin--Helmholtz phase. 
Note that $N_{\rm tot}$ does not always monotonically increase with $M_b$. Nevertheless, we find a positive correlation between $N(>\!1\,{\rm s})$ and $M_b$ for each EOS model. 
This is because the onset time of the PNS cooling simulation, $t_{\rm init}$, is different among the models (Table~\ref{tab:enumber}), and the models with a smaller $t_{\rm init}$ tend to have a higher number of events detected in the range $t_{\rm init} < t_{\rm pb} < 1$~s. 
The PNS mass dependence of the event number gets stronger at later times as does the EOS dependence.
In addition, a higher initial entropy leads to a larger event number as seen by comparing the models with the same PNS mass and EOS, but with different progenitor masses (e.g. model 162S15 and 162S40).
Since an initial model with a higher entropy is less bounded by thermal pressure, the gravitational energy emitted by neutrinos becomes larger. 
In fact, as shown in Figure~\ref{fig:ntot}, $N_{\rm tot}$ is approximately  proportional to $\Delta M_{\rm g}$, where $\Delta M_{\rm g}$ is the difference between the initial gravitational mass of the PNS and that of the cold neutron star with the same baryon mass. 
The entropy dependence is obvious in the early phase and becomes minor later on.

The present assumptions predict an event rate that is about 400 times higher than the yields for SN 1987A, \citep[whose distance was estimated at 51.2~kpc,][]{1991ApJ...380L..23P} with Kamiokande-II (2.14 kton), where $N(>\!10\,{\rm s})=2$ and $N(>\!20\,{\rm s})=0$ \citep{1987PhRvL..58.1490H}.
While the mass of SN 1987A's neutron star is still unknown, adopting the Togashi EOS implies a low mass.
This is because the event numbers are scaled to $N(>\!10\,{\rm s})=3.0$ and $N(>\!20\,{\rm s})=1.7$ for the model with $M_b=1.40M_\odot$, which has the lowest mass among our models. Nevertheless, it is not inconsistent with the SN 1987A data because ${\rm Poi}(2|3.0)=0.224$ and ${\rm Poi}(0|1.7)=0.183$, where ${\rm Poi}(N|m)$ is the Poisson probability of $N$ events for the mean value of $m$.

In the following, we focus on events in the period from the Kelvin--Helmholtz cooling.
About 100--2,000 events are expected for $t_{\rm pb}>20$~s. 
The neutrino signal at these late times is mainly determined by a relatively small number of parameters: the PNS mass, radius, and surface temperature \citep{2021PTEP.2021a3E01S}. 
This is the benefit of considering the long-term evolution of the neutrino emission and not its early phase, which is affected by other uncertain processes, such as mass accretion onto the PNS, convection processes, and hydrodynamical instabilities that may lead to shock revival.
Although the expected event number of neutrinos in the late phase is smaller than that in the early phase, the difference among models may be statistically significant as presented in the next sections.

\subsection{Backward Time Analysis of the Event Number} \label{sec:backward}
We use the reverse cumulative event number in Eq.~(\ref{eq:revcumev}) to estimate how long the neutrinos are detectable.
We define the time of the last expected event, $t_{\rm last}$, as $N(>\!t_{\rm last})=1$. 
As seen from Table~\ref{tab:enumber}, neutrino detection persists for about 40--170~s, where the duration depends on the PNS mass and the EOS. 
In particular, the EOS dependence is more drastic than the mass dependence and the Togashi EOS models have a much longer duration than the Shen EOS models. The duration of the LS220 EOS models is similar to that of the T+S EOS models when the PNS mass is the same.
While the T+S EOS has a longer cooling timescale in the Kelvin--Helmholtz phase, it has a steeper decline in event rate than the LS220 EOS after the transition to transparency.
For each EOS model the difference in $t_{\rm last}$ is subtle between the models with $(M_b, M_{\rm ZAMS})=(1.62M_\odot, 15M_\odot)$ and $(1.62M_\odot, 40M_\odot)$ and is attributed to the initial entropy. 
Note that since the event rate of supernova neutrinos at around $t_{\rm pb}=t_{\rm last}$ is still higher than the background rate of Super-Kamiokande after the spallation cut (defined below), our estimation does not suffer significantly from systematic errors on the background (see \S\ref{sec:bkg} for details).

\begin{figure*}[t]
\gridline{\fig{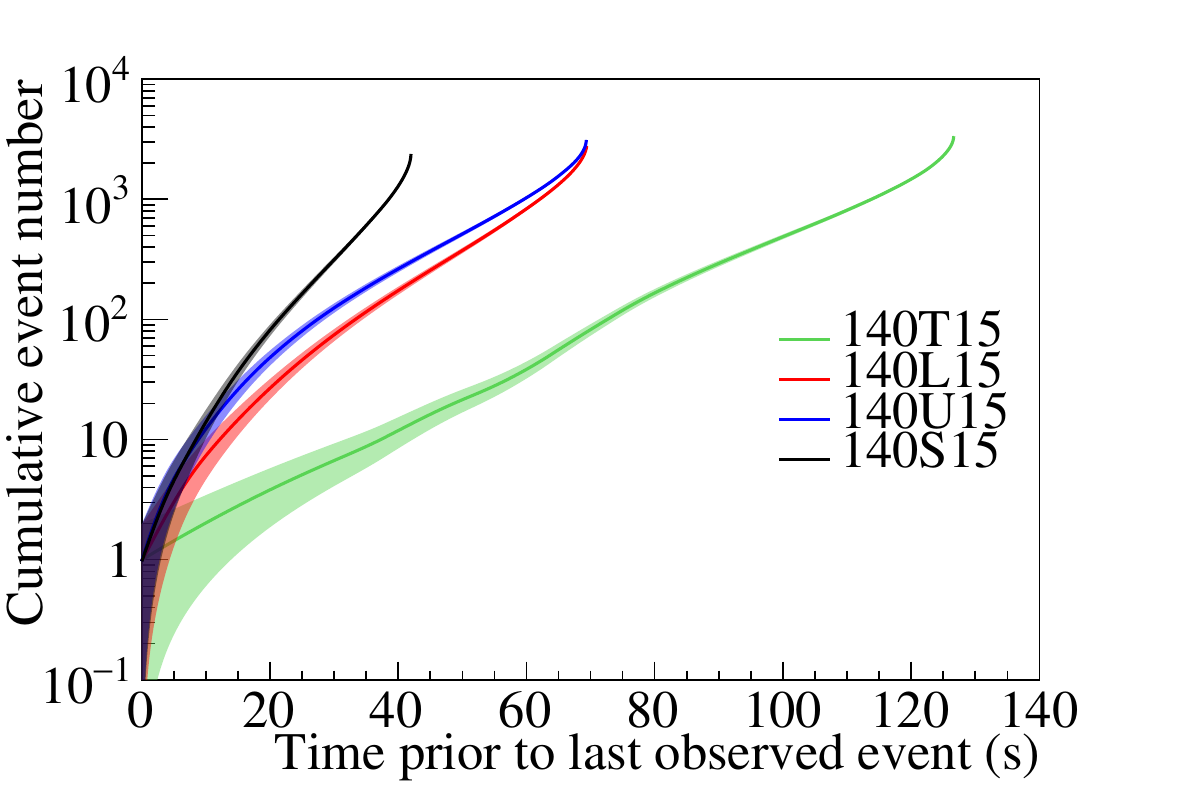}{0.495\textwidth}{(a) $M_b=1.40 M_\odot$}
          \fig{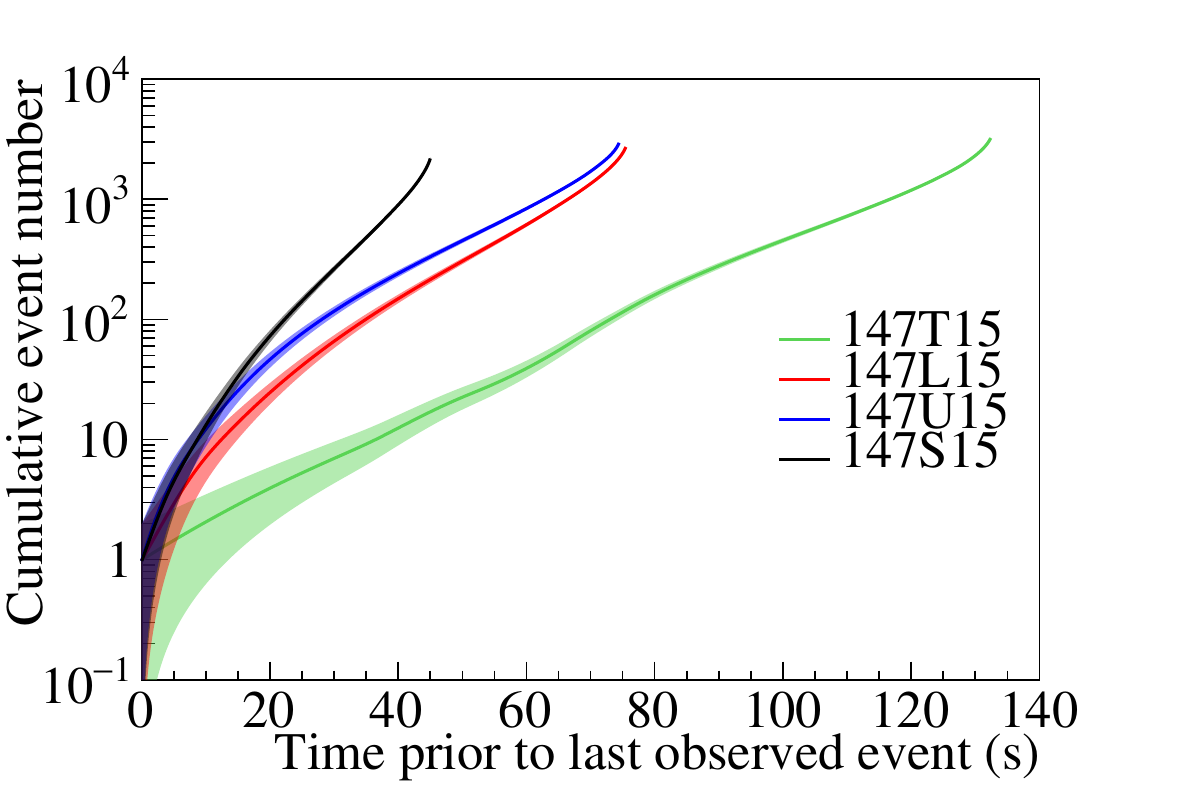}{0.495\textwidth}{(b) $M_b=1.47 M_\odot$}
          }
\gridline{\fig{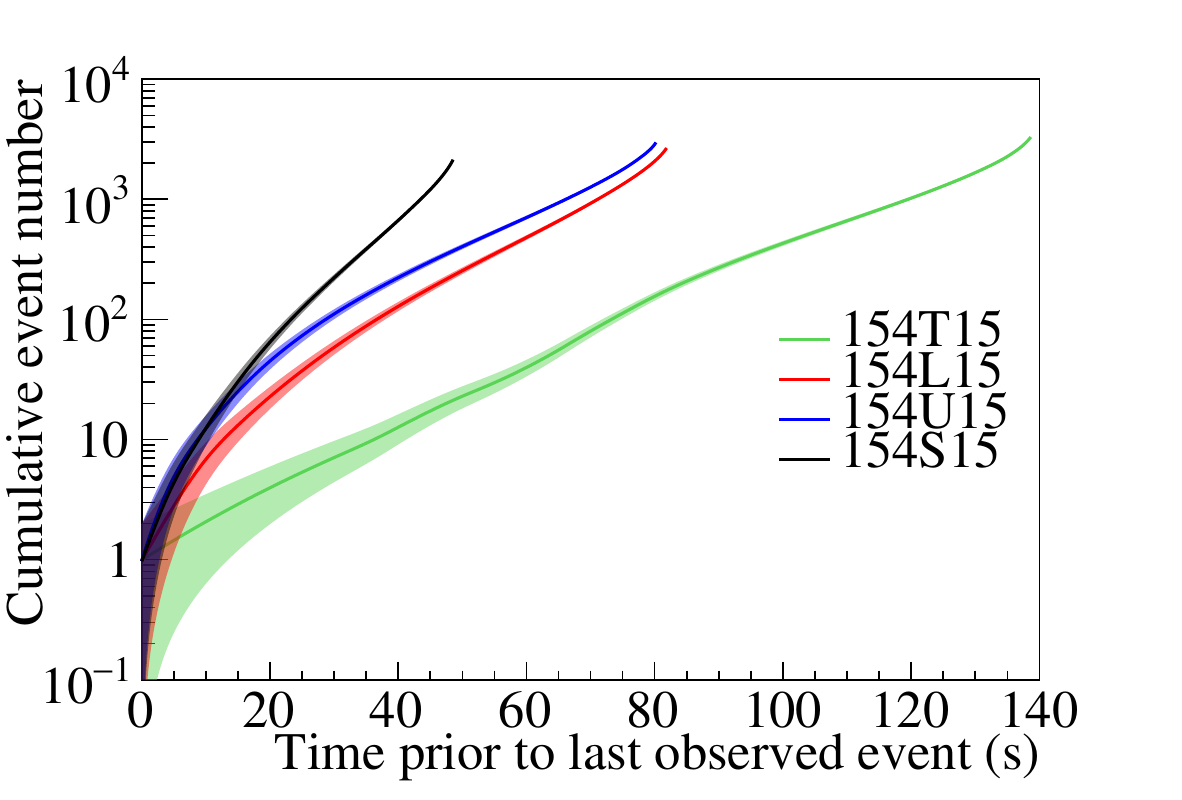}{0.495\textwidth}{(c) $M_b=1.54 M_\odot$}
          \fig{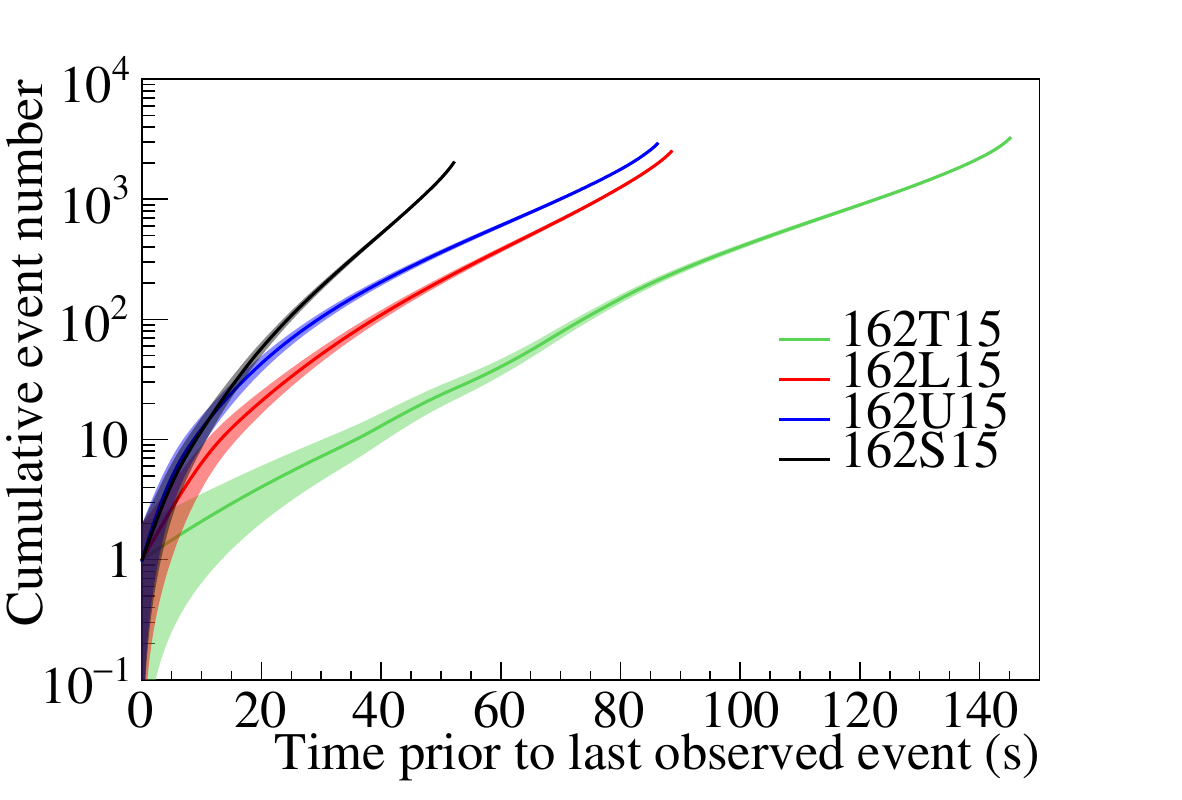}{0.495\textwidth}{(d) $M_b=1.62 M_\odot$}
          }
\caption{ EOS dependence of the cumulative event number for models with the $15 M_\odot$ progenitor. The lines correspond (from top to bottom) to the models with the Shen EOS, T+S EOS, LS220 EOS, and Togashi EOS. Here time is measured backward from the last expected event and the shaded region shows the variation in the prediction assuming statistical uncertainties.
\label{fig:bw15}}
\end{figure*}

\begin{figure*}[t]
\gridline{\fig{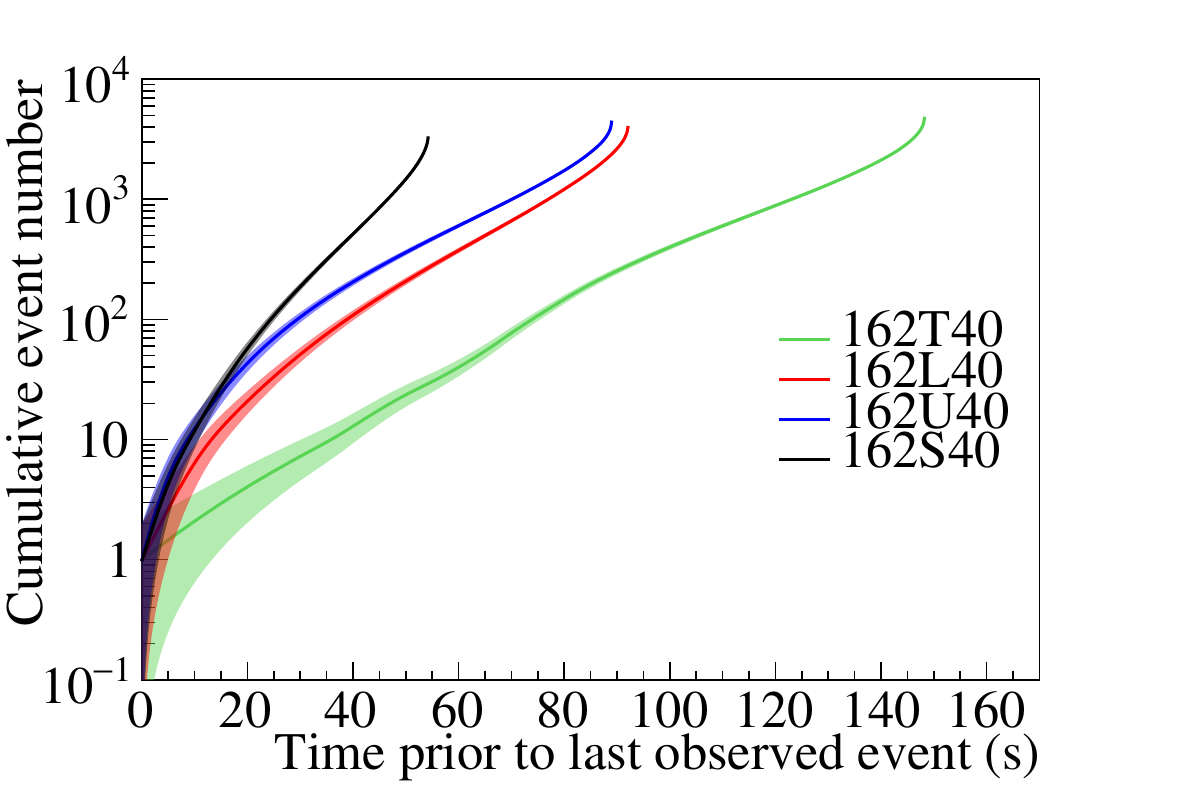}{0.495\textwidth}{(a) $M_b=1.62 M_\odot$}
          \fig{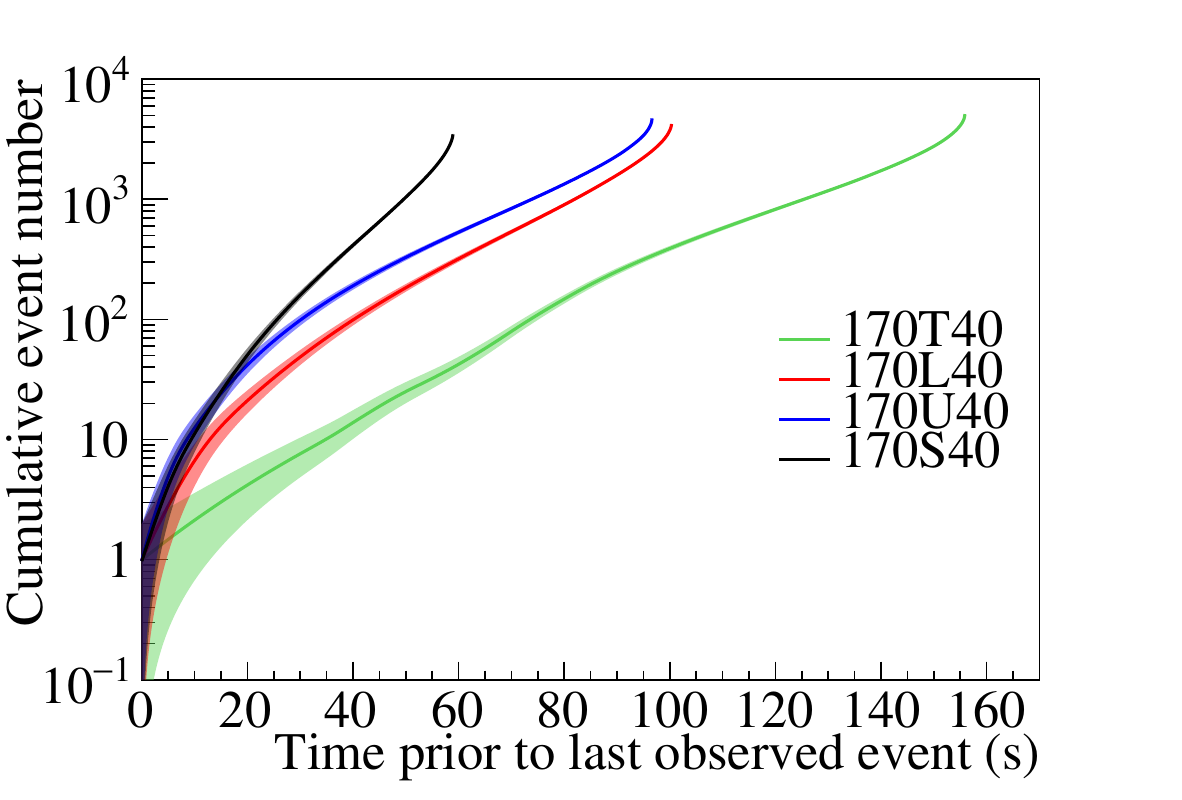}{0.495\textwidth}{(b) $M_b=1.70 M_\odot$}
          }
\gridline{\fig{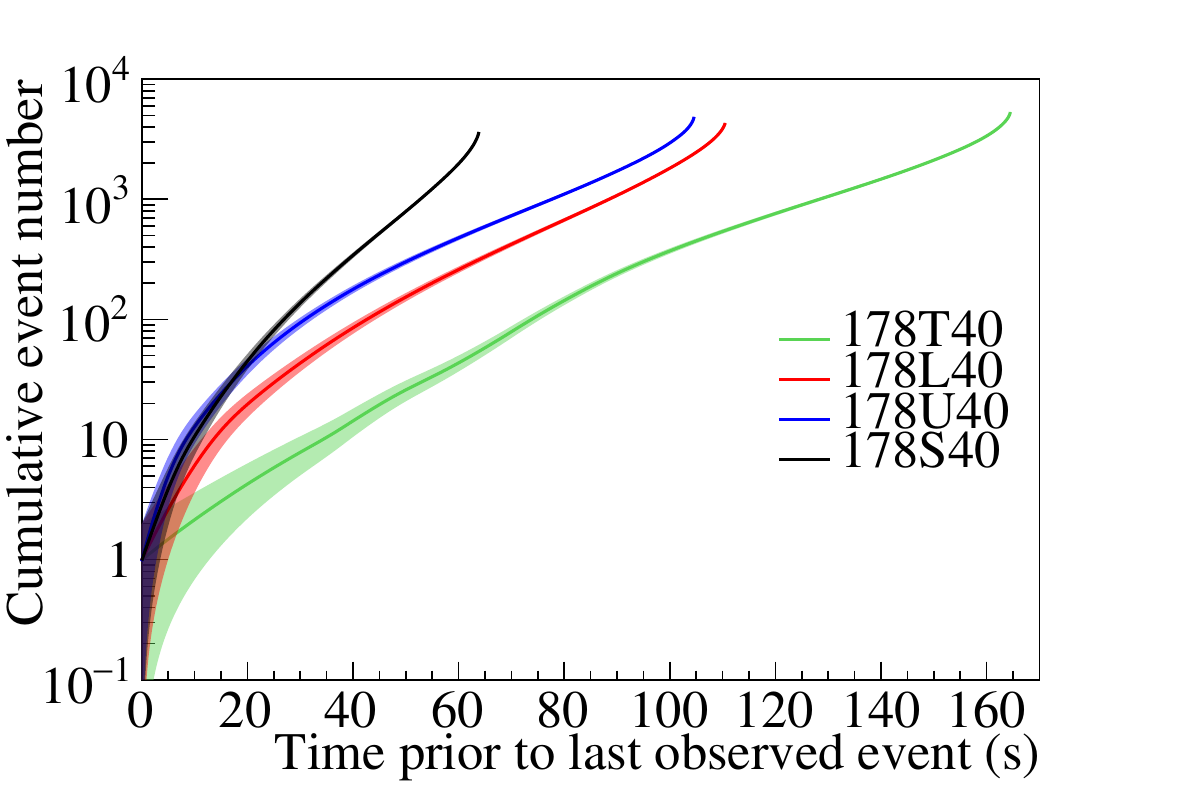}{0.495\textwidth}{(c) $M_b=1.78 M_\odot$}
          \fig{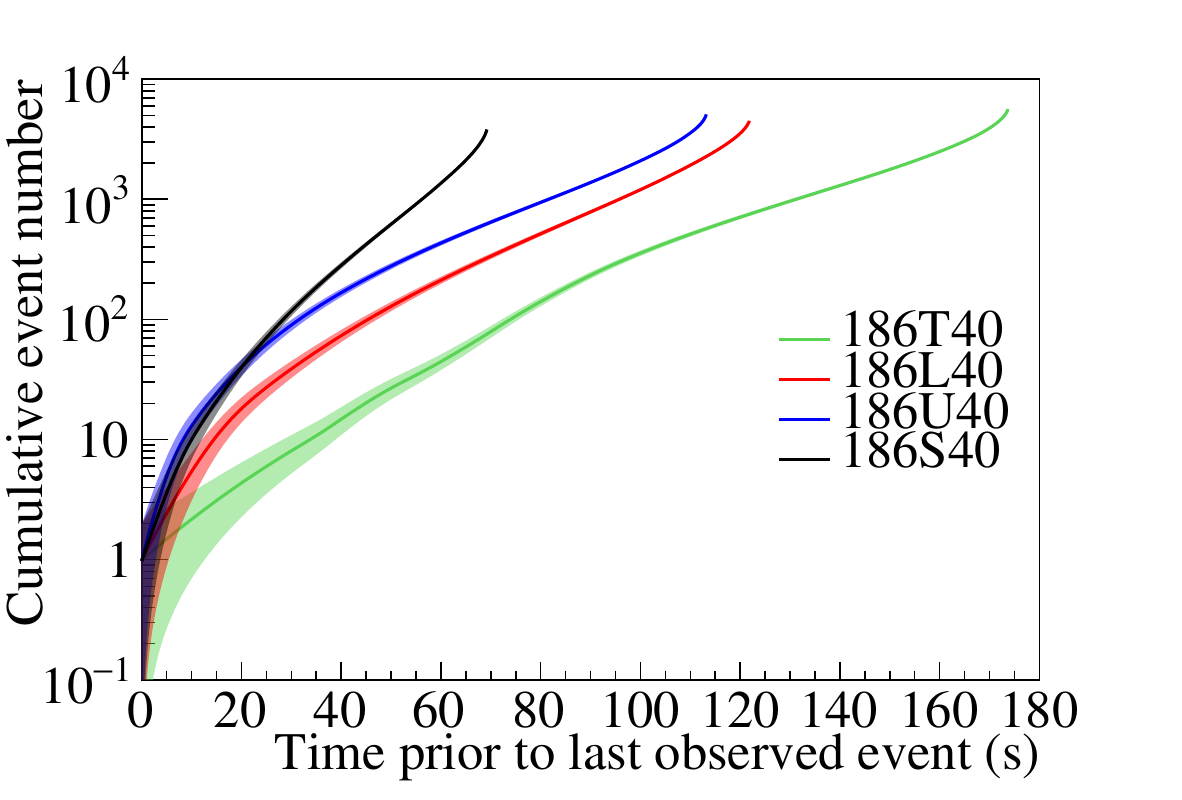}{0.495\textwidth}{(d) $M_b=1.86 M_\odot$}
          }
\caption{Same as Figure~\ref{fig:bw15} but for models with the $40 M_\odot$ progenitor.
\label{fig:bw40}}
\end{figure*}

In order to explore the properties of the PNS, representing the cumulative event number as a function of time measured backward from $t_{\rm last}$ was proposed in Paper I.
The backward time analysis of the event number is useful to disentangle PNS properties that are washed out in the early phase by uncertainties such as that of the initial entropy profile. 
In Figures \ref{fig:bw15} and \ref{fig:bw40} the backward cumulative event numbers with Poisson statistical uncertainties are compared for different EOS models.
We see that the cumulative event numbers of the Togashi EOS models have a shallow slope while those of the Shen EOS and T+S EOS models show a steep gradient around the time origin. 
The event rate of the Togashi EOS models varies slowly even in the late phase because the surface temperature of the PNS is high due to its abundance of heavy nuclei. On the other hand, models with the Shen EOS and the T+S EOS are similar in the behavior of their backward cumulative event number around the time origin.
This is consistent with the fact that both EOSs share the low-density region where heavy nuclei reside.
In addition, the difference in the time evolution of the cumulative event number is significant between models with the LS220 EOS and T+S EOS though they have similar $t_{\rm last}$ values.

\begin{figure*}[t]
\gridline{\fig{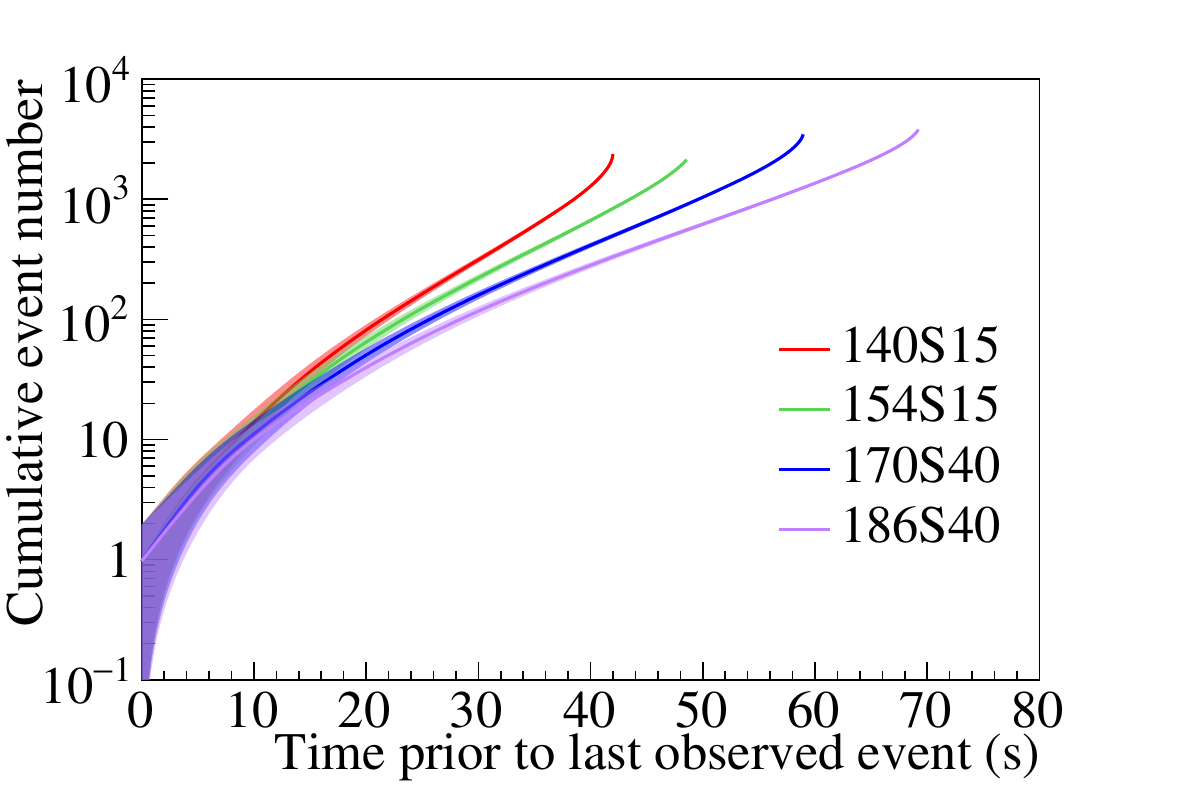}{0.495\textwidth}{(a) Shen EOS}
          \fig{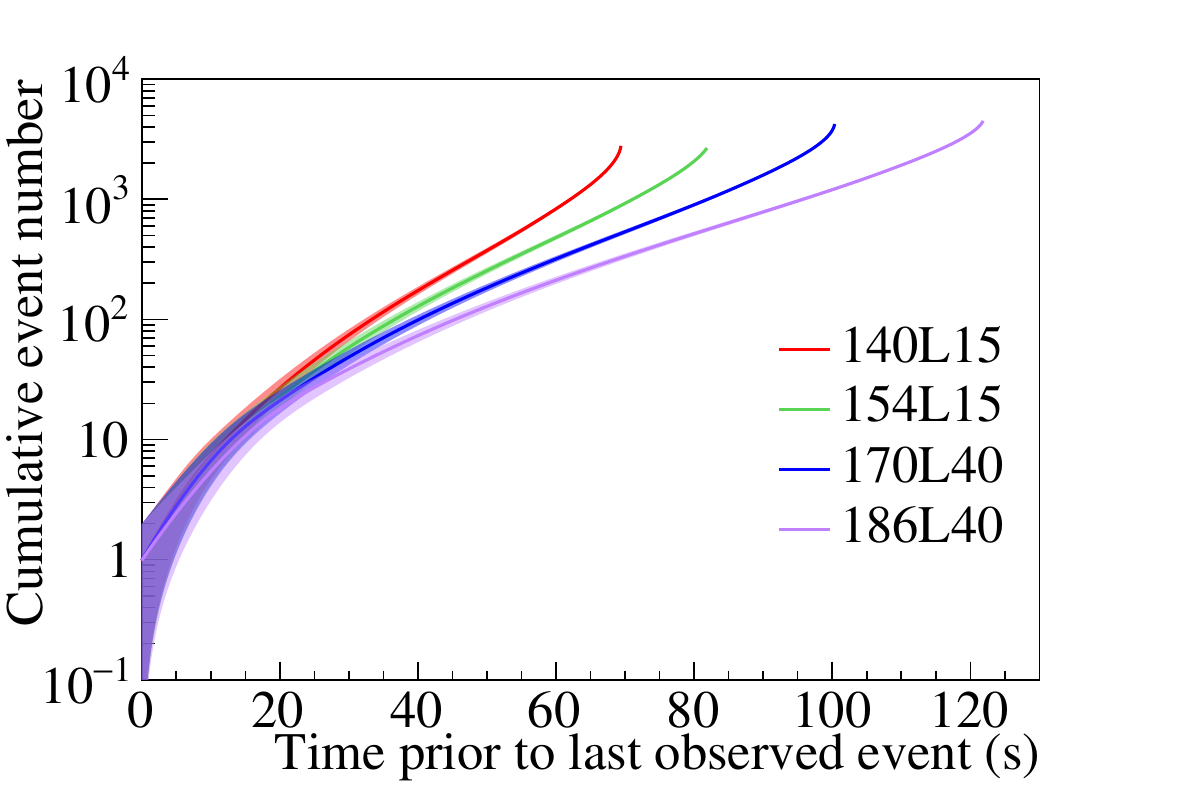}{0.495\textwidth}{(b) LS220 EOS}
          }
\gridline{\fig{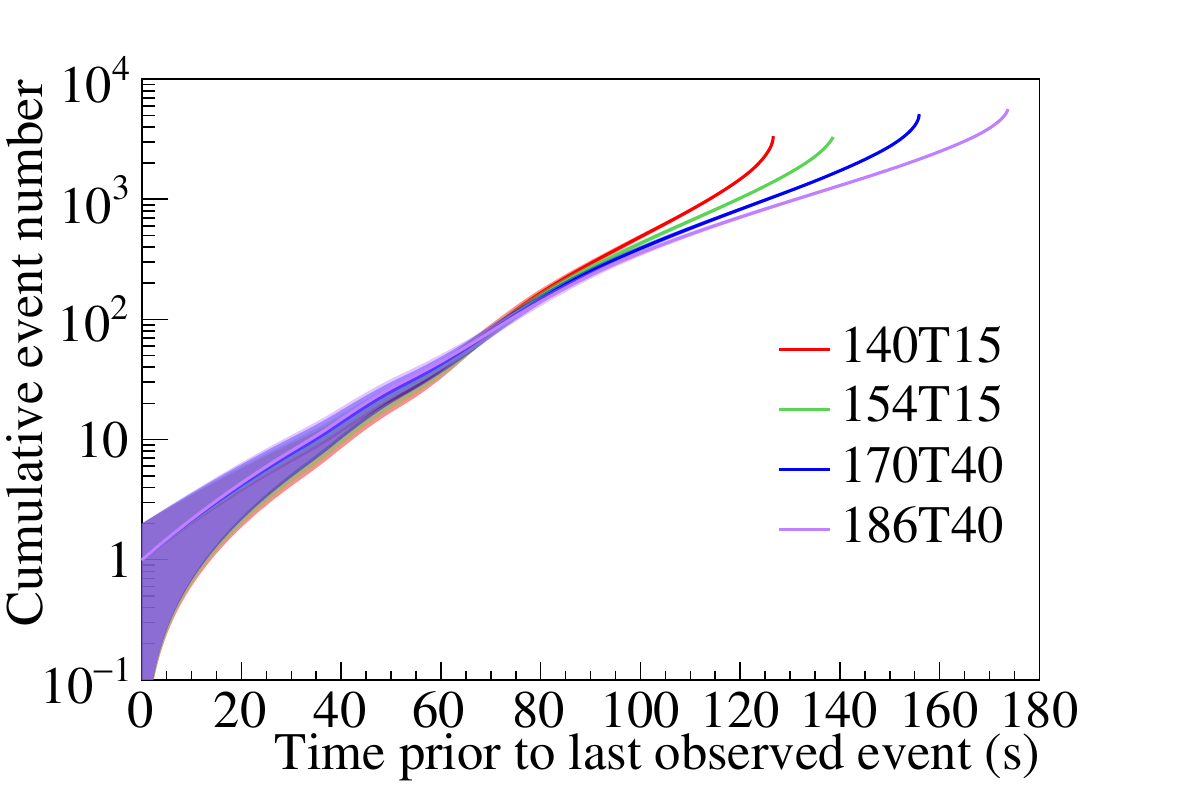}{0.495\textwidth}{(c) Togashi EOS}
          \fig{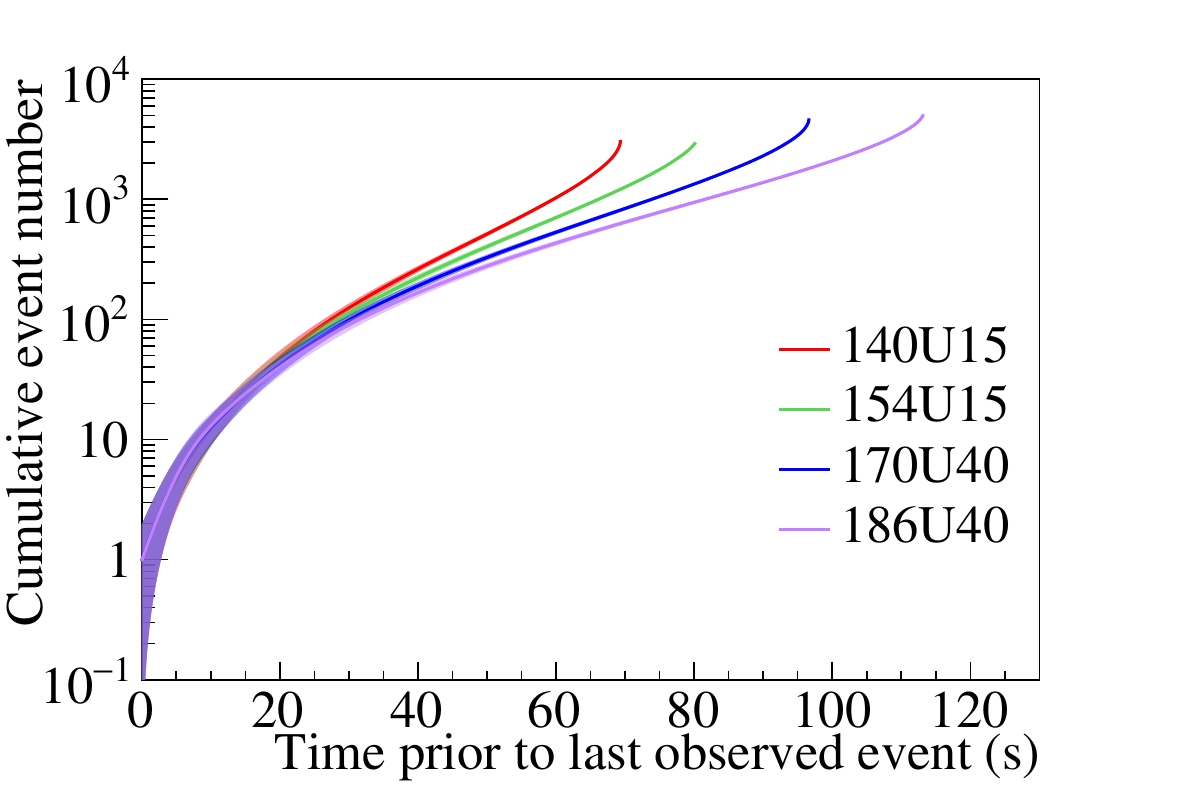}{0.495\textwidth}{(d) T+S EOS}
          }
\caption{Same as Figure~\ref{fig:bw15} except that PNS models with different masses are compared in each panel, where the lines correspond, from top to bottom, to  models with $(M_b, M_{\rm ZAMS})=(1.40M_\odot, 15M_\odot)$, $(1.54M_\odot, 15M_\odot)$, $(1.70M_\odot, 40M_\odot)$, and $(1.86M_\odot, 40M_\odot)$.
\label{fig:bwEOS}}
\end{figure*}

The PNS mass dependence of the backward cumulative event number is illustrated in Figure~\ref{fig:bwEOS}. 
Bands with different PNS masses are well separated, especially in the early phase. In contrast, the EOS dependence is still significant in the late phase as stated above. 
In the event of actual supernova neutrino detection, theoretical predictions of the cumulative event number evolution will be useful as immediately comparable templates for extracting information about the PNS mass as well as the EOS.
In Paper I we have reported that the neutrino oscillation effect has no influence on the backward time analysis, particularly in the late phase.

\subsection{Backward Time Analysis of the Event Energy} \label{sec:bwene}

The average energy of recoil positrons provides complementary information to the cumulative event number. 
Measurement of the average energy can be compared with theoretical models even if the distance to the supernova is unknown. 
Unfortunately, reconstructing the time evolution of the average energy is more difficult than that of the cumulative event number due to statistical uncertainties. Since the incident neutrinos have a distribution of energies, a meaningful estimation of the average energy requires a reasonable number of events. 
While we can reduce the uncertainty in the average energy by averaging over a larger number of events, wider time bins are required to do so.
In the present work we propose a strategy to follow the evolution of the average energies of recoil positrons and incident neutrinos, performing a
MC simulation to evaluate uncertainties on the average event energies.

\begin{figure*}[t]
\gridline{\fig{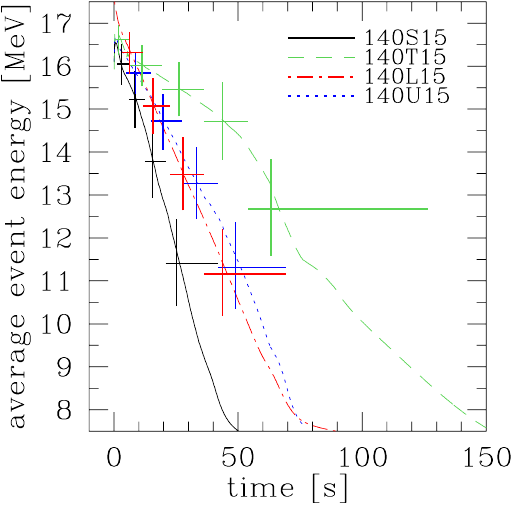}{0.249\textwidth}{(a) $M_b=1.40 M_\odot$, $M_{\rm ZAMS}=15 M_\odot$}
          \fig{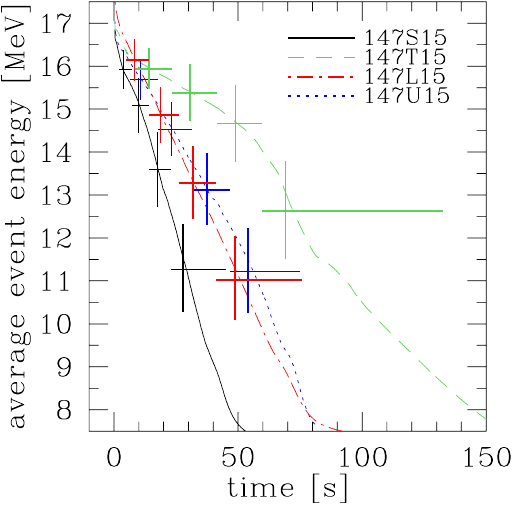}{0.249\textwidth}{(b) $M_b=1.47 M_\odot$, $M_{\rm ZAMS}=15 M_\odot$}
          \fig{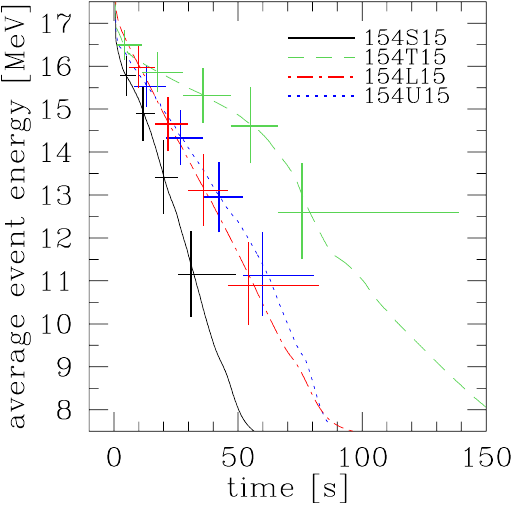}{0.249\textwidth}{(c) $M_b=1.54 M_\odot$, $M_{\rm ZAMS}=15 M_\odot$}
          \fig{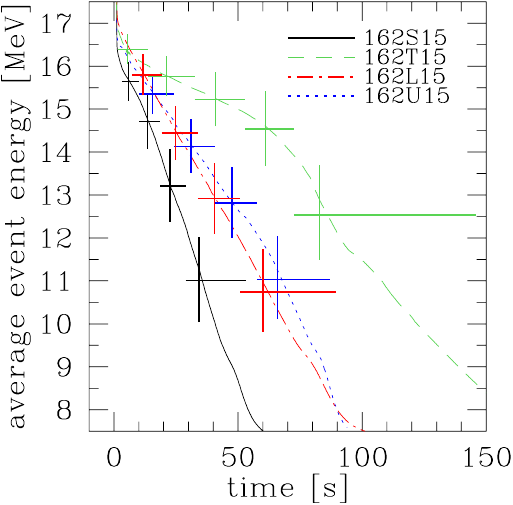}{0.249\textwidth}{(d) $M_b=1.62 M_\odot$, $M_{\rm ZAMS}=15 M_\odot$}
          }
\gridline{\fig{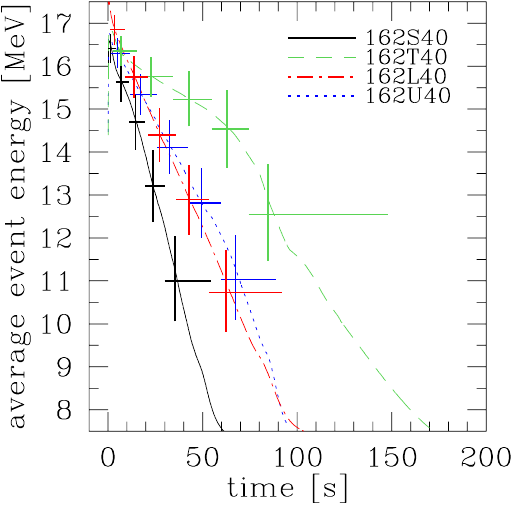}{0.249\textwidth}{(e) $M_b=1.62 M_\odot$, $M_{\rm ZAMS}=40 M_\odot$}
          \fig{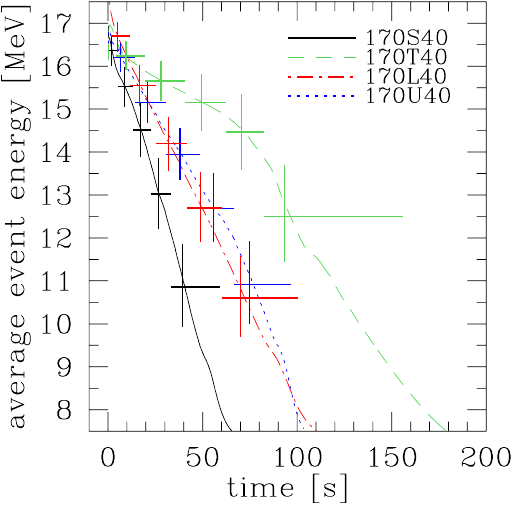}{0.249\textwidth}{(f) $M_b=1.70 M_\odot$, $M_{\rm ZAMS}=40 M_\odot$}
          \fig{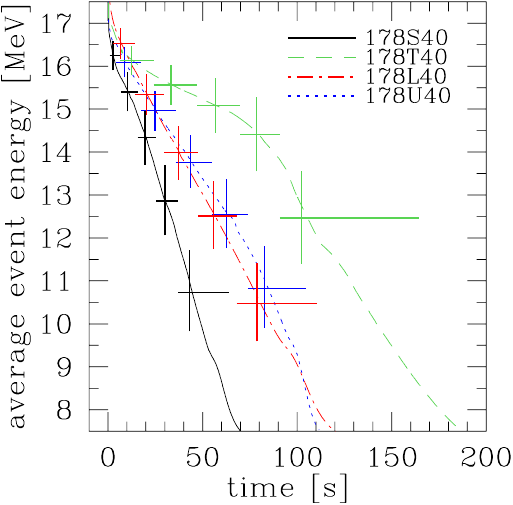}{0.249\textwidth}{(g) $M_b=1.78 M_\odot$, $M_{\rm ZAMS}=40 M_\odot$}
          \fig{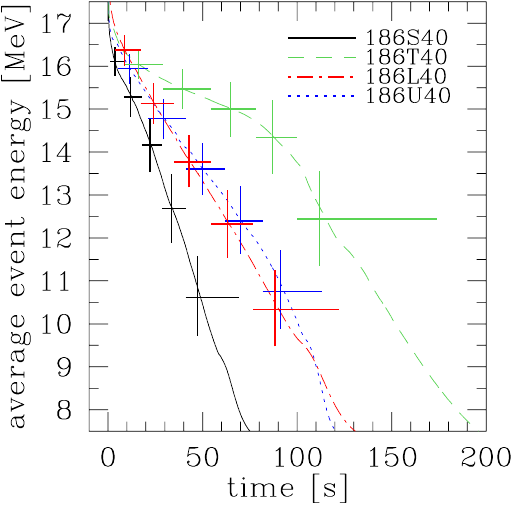}{0.249\textwidth}{(h) $M_b=1.86 M_\odot$, $M_{\rm ZAMS}=40 M_\odot$}
          }
\caption{Time evolution of the average IBD event energy with errors in the late phase of a supernova at 10~kpc. 
PNS models with different EOSs are compared in each panel, where solid (black), dashed (green), dotted (blue), and dot-dashed (red) lines correspond to the models with the Shen, the Togashi, the T+S, and the LS220 EOSs, respectively. Backward in time, the individual bins contain 100, 200, 400, 800, and 1600 (if the fifth bin exists) events for each model. 
The length of the error bars corresponds to the 95\% coverage of the average event energy obtained from MC simulations and the width of the error bars indicates the chosen time interval. 
Horizontal and vertical error bars cross at the time of the data median in each bin.
\label{fig:averrm}}
\end{figure*}

\begin{figure*}[t]
\gridline{\fig{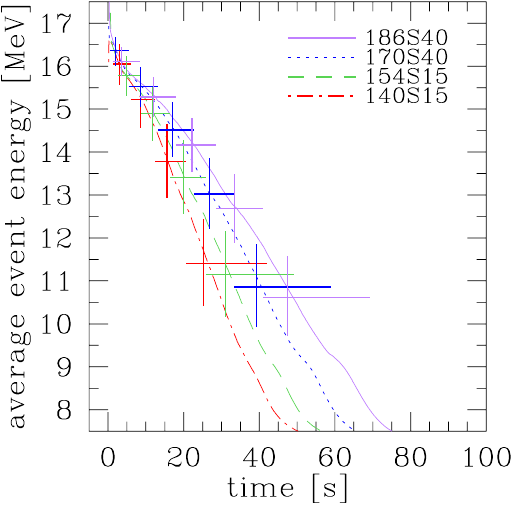}{0.249\textwidth}{(a) Shen EOS}
          \fig{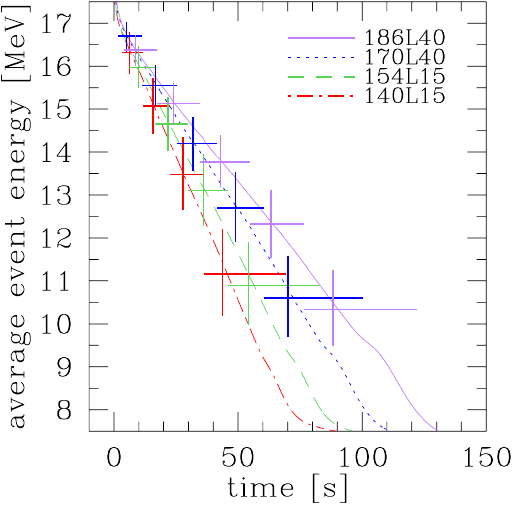}{0.249\textwidth}{(b) LS220 EOS}
          \fig{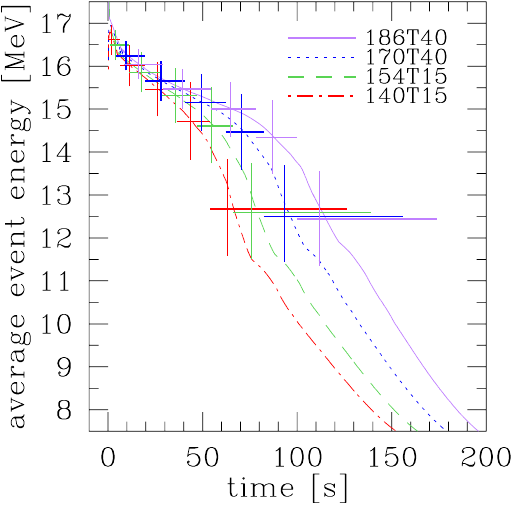}{0.249\textwidth}{(c) Togashi EOS}
          \fig{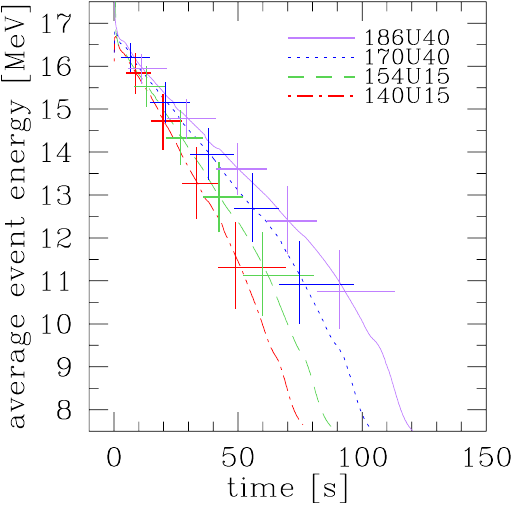}{0.249\textwidth}{(d) T+S EOS}
          }
\caption{Same as Figure~\ref{fig:averrm} except that the PNS models with different masses are compared in each panel, where dot-dashed (red), dashed (green), dotted (blue), and solid (purple) lines correspond to the models with $(M_b, M_{\rm ZAMS})=(1.40M_\odot, 15M_\odot)$, $(1.54M_\odot, 15M_\odot)$, $(1.70M_\odot, 40M_\odot)$, and $(1.86M_\odot, 40M_\odot)$, respectively.
\label{fig:averre}}
\end{figure*}

To estimate the average energy in our backward time analysis, the last 100 events are binned into a single time bin for evaluation.
Subsequent bins are chosen such that the number of events contained therein increases exponentially going backward in time; the last bin has 100 events, the second to last bin has 200 events, the third to last bin has 400 events, and so on. The uncertainty on the average energy evaluated in this way is shown in Figures~\ref{fig:averrm} and \ref{fig:averre}.
The width of the error bars denotes the chosen time intervals satisfying the above criteria and therefore changes with the expected number of events in each MC simulation.
Note that this shifts the boundary between the last two bins by at most $\pm 3$~s from simulation to simulation.
This approach has the advantage of ensuring a minimum statistical uncertainty on the measured average in each bin.

In Figure~\ref{fig:averrm} the average energy of recoil positrons and its uncertainty are compared for different EOS models. 
For models with the Shen EOS, the Togashi EOS, and the T+S EOS, the difference in the evolution of the average energy is significant. 
In particular, the Togashi EOS model can be distinguished by its high event energy relative to those of other models.
We recall that the T+S EOS is different from the Shen EOS in the high-density region and different from the Togashi EOS in the low-density region.
However, this analysis is incapable of distinguishing the LS220 EOS from the T+S EOS for a supernova distance of 10~kpc.
Figure~\ref{fig:averre} compares the models with different PNS masses. 
While the models with $M_b=1.40M_\odot$ and $1.86M_\odot$ are separated for each EOS, the backward time analysis for the average energy is not as useful for determining the PNS mass in comparison to the cumulative event number analysis.
Nevertheless, the event energy is advantageous in that it does not in principle depend on distance to the supernova.

\begin{figure}[t!]
\epsscale{0.5}
\plotone{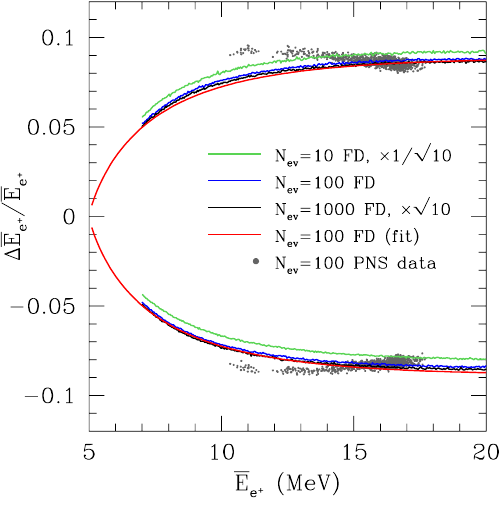}
\caption{Statistical error on the measured average energy of recoil positrons as a function of the true value. Red lines show the approximation in Eq.~(\ref{eq:FDbar_E_e_err}) and other lines are obtained from MC simulations based on a Fermi--Dirac distribution. Grey dots denote the MC results with $N_{\rm ev}=100$ based on the PNS cooling simulations.
\label{fig:err100}}
\end{figure}

\begin{figure}[t!]
\epsscale{0.5}
\plotone{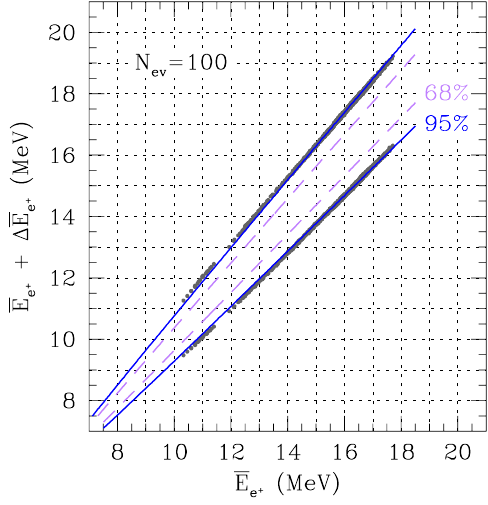}
\caption{Same as Figure~\ref{fig:err100} except that $\bar E_{e^+}+\Delta \bar E_{e^+}$ is plotted as a function of $\bar E_{e^+}$. Solid and dashed lines represent, respectively, the 95\% and 68\% C.L. expected ranges from MC simulations based on the Fermi--Dirac distribution for the case with $N_{\rm ev}=100$. The grey dots denote the MC results based on the PNS cooling simulations.
\label{fig:err100inv}}
\end{figure}

Here we discuss statistical errors on the measured average energy of recoil positrons from IBD interactions. 
For this purpose we consider neutrinos with a phase space occupation function given by a Fermi--Dirac function with no chemical potential $f_{\rm FD}(E_\nu)=1/(1+e^{E_\nu/k_B T_\nu})$, where $k_B$ is the Boltzmann constant and $T_\nu$ is the neutrino temperature.
The neutrino energy distribution function is then proportional to $E_\nu^2f_{\rm FD}(E_\nu)$.
We assume that the IBD cross section is $\sigma(E_\nu)\propto E_\nu^2$, neglecting the mass difference between neutrons and protons (i.e., $E_{e^+} \simeq E_\nu$) for simplicity. 
Since the energy spectrum of the recoil positrons is obtained as the product of the cross section and the neutrino distribution, it is proportional to $E_{e^+}^4f_{\rm FD}(E_{e^+})$. Thus, the average energy of recoil positrons can be written as
\begin{equation}
    \bar E_{e^+}=
    \frac
    {\displaystyle\int_{E_{\rm th}}^{\infty}E_{e^+}^5f_{\rm FD}(E_{e^+})dE_{e^+}}{\displaystyle\int_{E_{\rm th}}^{\infty}E_{e^+}^4f_{\rm FD}(E_{e^+})dE_{e^+}},
    \label{eq:FDbar_E_e}
\end{equation}
and is evaluated in Table~\ref{tab:positron_energy} for different neutrino temperatures and threshold energies (Paper I). 
Under these assumptions we estimate the statistical error on the measured average energy of recoil positrons by carrying out MC simulations. In Table~\ref{tab:positron_energy}, we show the 95\% confidence level (C.L.) expected range for the case with an observed event number of $N_{\rm ev}=100$. 
We recognize that the error, $\Delta \bar E_{e^+}$, is about $\pm 1$~MeV or $\Delta \bar E_{e^+}/\bar E_{e^+}\sim\pm 8$--9\% for the range $10~{\rm MeV} \leq \bar E_{e^+} \leq 18~{\rm MeV}$, which is consistent with models based on the PNS cooling simulations as shown in Figure~\ref{fig:err100}. 
Furthermore, for the range covered in Table~\ref{tab:positron_energy}, we find that the error is well approximated by 
\begin{equation}
    \Delta \bar E_{e^+}=\pm\frac{1}{\sqrt{N_{\rm ev}}}\left(\frac{\bar E_{e^+}-E_{\rm th}}{\rm MeV}\right)^{0.8}\left(0.1E_{\rm th}+1.5~{\rm MeV}\right),
    \label{eq:FDbar_E_e_err}
\end{equation}
provided that $N_{\rm ev} \geq 100$. This approximation is compared with the MC results in Figure~\ref{fig:err100}.
In Figure~\ref{fig:err100inv} we show $\bar E_{e^+}+\Delta \bar E_{e^+}$ as a function of $\bar E_{e^+}$, which is the true value of the average energy of IBD events.
While the average energy actually measured in the observation includes the statistical error, we will be able to estimate its true value using this figure.

\begin{table}[]
\centering
    \caption{Average energy of positrons from IBD reactions evaluated by Eq.~(\ref{eq:FDbar_E_e}) and its 95\% C.L. expected range for an observation of  $N_{\rm ev}=100$ obtained from MC simulations.}
    \begin{tabular}{lrrr}
\hline \hline
$k_BT_\nu$ & \multicolumn{3}{c}{$\bar E_{e^+}$ (MeV)} \\ \cline{2-4}
(MeV) & $E_{\rm th}=3$ MeV & $E_{\rm th}=5$ MeV & $E_{\rm th}=7$ MeV \\
         \hline
         1 & $5.63_{-0.39}^{+0.42}$ & $7.00_{-0.33}^{+0.36}$ & $8.69_{-0.30}^{+0.33}$ \\
         1.5 & $7.85_{-0.62}^{+0.67}$ & $8.73_{-0.57}^{+0.61}$ & $10.10_{-0.51}^{+0.56}$ \\
         2 & $10.25_{-0.86}^{+0.89}$ & $10.80_{-0.81}^{+0.85}$ & $11.84_{-0.75}^{+0.80}$ \\
         2.5 & $12.73_{-1.07}^{+1.12}$ & $13.08_{-1.03}^{+1.09}$ & $13.84_{-0.98}^{+1.04}$ \\
         3 & $15.24_{-1.29}^{+1.36}$ & $15.46_{-1.27}^{+1.34}$ & $16.02_{-1.22}^{+1.29}$ \\
         3.5 & $17.76_{-1.51}^{+1.59}$ & $17.90_{-1.49}^{+1.57}$ & $18.31_{-1.45}^{+1.53}$ \\
         4 & $20.26_{-1.73}^{+1.80}$ & $20.37_{-1.72}^{+1.80}$ & $20.67_{-1.69}^{+1.77}$ \\
         \hline
    \end{tabular}
    \label{tab:positron_energy}
\end{table}


We now turn to the average energy of supernova neutrinos estimated from the measured average energy of recoil positrons. 
As a first step we again adopt the assumptions described above (the Fermi--Dirac distribution for neutrinos, $E_{e^+} \simeq E_\nu$ for the positron energy, and $\sigma(E_\nu)\propto E_\nu^2$ for the IBD cross section). 
The average neutrino energy is then given as $\bar E_\nu = 3.15k_B T_\nu$. 
Moreover, if the average energy is much higher than the threshold energy of positron detection, we can assume $E_{\rm th}=0$ and obtain $\bar E_{e^+} = 5.07k_B T_\nu=1.61\bar E_\nu$.
Otherwise the average positron energy will converge to $E_{\rm th}$ as $\bar E_\nu \to 0$. 
So as to connect the two extremes we approximate the average neutrino energy as $\left. \bar E_\nu \simeq\sqrt{\bar E_{e^+}^2 - E_{\rm th}^2} \middle/ 1.61 \right.$ by adopting the following replacement;
\begin{equation}
    \bar E_{e^+} \to \sqrt{\bar E_{e^+}^2 - E_{\rm th}^2}.
    \label{eq:recstr_approx}
\end{equation}
Note that the values obtained from this approximation differ from those in Table~\ref{tab:positron_energy} by at most 5\%. 
In the second step we consider corrections to the positron energy and IBD cross section up to first order in $1/M$, where $M$ is the nucleon mass:
\begin{equation}
    \sigma(E_\nu)\propto E_\nu^2 \left( 1 + A\frac{E_\nu}{M}\right),
\label{eq:1st_ep}
\end{equation}
\begin{equation}
    E_{e^+} \simeq E_\nu \left( 1 + B\frac{E_\nu}{M}\right),
\label{eq:1st_cs}
\end{equation}
with coefficients $A$ and $B$. 
Note that we keep the assumption that the mass difference between neutrons and protons is negligible. 
If $E_{\rm th}=0$ is assumed the average energy of recoil positrons is given as
\begin{eqnarray}
    \bar E_{e^+} & \simeq &
    \frac
    {\int_{0}^{\infty}E_{\nu}^5\left(1 + A\frac{E_\nu}{M}\right)\left(1 + B\frac{E_\nu}{M}\right)f_{\rm FD}(E_{\nu})dE_{\nu}}{\int_{0}^{\infty}E_{\nu}^4\left(1 + A\frac{E_\nu}{M}\right)f_{\rm FD}(E_{\nu})dE_{\nu}} \nonumber \\
    & \simeq & 5.07k_B T_\nu - A\frac{(5.07 k_B T_\nu)^2}{M} + (A+B)\frac{30.6(k_B T_\nu)^2}{M} \nonumber \\
    & \simeq & 1.61\bar E_\nu \left(1-4\frac{\bar E_\nu}{M}\right),
    \label{eq:FDbar_E_e_1st}
\end{eqnarray}
where $A\simeq -7$ and $B\simeq -0.966$ \citep{1999PhRvD..60e3003V}\footnote{We note that the $1/M$ corrections of Eqs.~(\ref{eq:1st_ep}) and (\ref{eq:1st_cs}) are not the full results of \citet{1999PhRvD..60e3003V}, which provides other corrections at different orders.} and $\bar E_\nu = 3.15k_B T_\nu$ is used. 
For the case where the average neutrino energy is comparable to $E_{\rm th}$, we again make the replacement in Eq.~(\ref{eq:recstr_approx}) and obtain
\begin{equation}
    \bar E_{\nu} \simeq \frac{\sqrt{\bar E_{e^+}^2 - E_{\rm th}^2}}{1.61}+\frac{4}{M} \left(\frac{\bar E_{e^+}^2 - E_{\rm th}^2}{1.61^2}\right),
    \label{eq:recstr_enu1st}
\end{equation}
where Eq.~(\ref{eq:FDbar_E_e_1st}) is solved for $\bar E_\nu$.

\begin{figure}[t!]
\plottwo{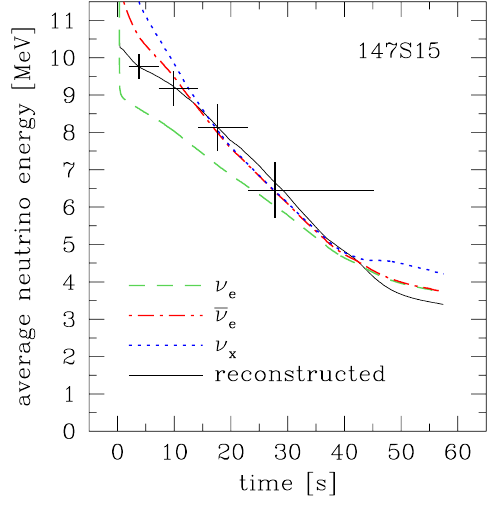}{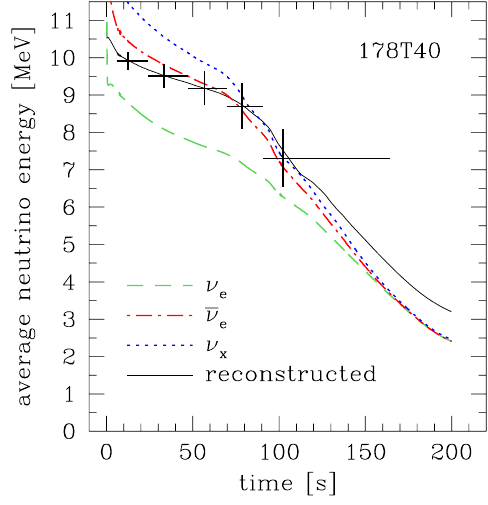}
\caption{Reconstructed time evolution of the average neutrino energy for models 147S15 (left) and 178T40 (right). Error bars and thin solid lines represent the reconstructed average energy of $\bar \nu_e$ using Eq.~(\ref{eq:recstr_enu1st}) from the expected average energy of IBD events shown in Figure~\ref{fig:averrm}. The error bars are drawn in the same manner as in Figure~\ref{fig:averrm}. 
Dashed, dot-dashed, and dotted lines correspond to the original average energies of $\nu_e$, $\bar \nu_e$, and $\nu_x$, respectively, obtained from PNS cooling simulations.
\label{fig:stat_enu}}
\end{figure}

In Figure~\ref{fig:stat_enu} we show examples of the reconstructed average neutrino energy using Eq.~(\ref{eq:recstr_enu1st}). 
Comparing it with the original average energy of $\bar \nu_e$ obtained from the PNS cooling simulations, we can see that this approximation works well for IBD events with $\bar E_{e^+} \lesssim 15$~MeV or reconstructed average neutrino energy $\bar E_{\nu} \lesssim 9$~MeV.
Without $1/M$ corrections this changes to $\bar E_{\nu} \lesssim 8$~MeV.
This is also the case for other models not shown in Figure~\ref{fig:stat_enu}.
In this regime systematic errors from the reconstruction, which are mainly attributed to the spectral difference between the Fermi-–Dirac distribution and simulation results of PNS cooling, are comparable or less than the statistical error. 
Thus, in this case it is hard to extract information beyond the average neutrino energy such as parameters controlling spectral deformations \citep[e.g., the pinching parameter in][]{2012PhRvD..86l5031T}.
Note that in the present analysis we focus on the IBD reaction of $\bar \nu_e$ and we do not take into account other reactions, which can be rejected in principle with neutron tagging. 
The average energies of $\nu_e$ and $\nu_x$ are also plotted in Figure~\ref{fig:stat_enu} and the difference in the average energies of $\bar\nu_e$ and $\nu_x$ at later times are confirmed to be smaller than the statistical error of our analysis. 
On the contrary, differences between flavors are larger than the errors during the early phase. Although the effects of neutrino oscillation are not considered in the present work, oscillations that exchange $\bar\nu_e$ and $\nu_x$ should be taken into account to investigate the average positron energy in this phase.

\section{Discussion}
\subsection{Data Analysis Strategy} \label{sec:dataanl}
In Paper I we proposed a data analysis strategy for the next Galactic supernova neutrino burst. In the present study we have improved the method to extract the average neutrino energy which thereby indicates the temperature of the PNS. This is summarized as follows:
\begin{enumerate}[(i)]
\item Sort observed events in reverse time order and divide them into time bins such that number of events in a bin increases exponentially, for instance from 100, 200, 400 events and so on. 
\item Calculate the average energy of the events and its error via  Eq.~(\ref{eq:FDbar_E_e_err}) for each bin.
\item Using Eq.~(\ref{eq:recstr_enu1st}) reconstruct the time sequence of the average energy of the original supernova neutrinos and its error.
\end{enumerate}
We can thereby identify the properties of the PNS born in the next Galactic supernova by comparing the time profile of the average neutrino energy obtained from this method with profiles from theoretical models such as that in this paper and the analytic solutions derived in \citet{2021PTEP.2021a3E01S}. 
In particular, the temperature evolution of the PNS, which is inferred from the neutrino average energy, will be useful to study the EOS of the inhomogeneous phase at subnuclear densities. 
We will apply this method to our integrated supernova neutrino analysis framework \citep{2020arXiv201016254M} in the near future.

\subsection{Detector Background}
\label{sec:bkg}
For a real observation some background events will be recorded in the detector. The dominant background sources are expected to be from radioactive isotopes created by spallation processes or from impurities in the detector's materials. 
Here spallation refers light radioactive isotopes created from the break up of oxygen nuclei by high energy cosmic-ray muons traversing the detector.
Though the muon rate can be high, for instance $\sim$2~Hz at Super-Kamiokande, 
spallation events are likely to be both spatially and temporally correlated with the track of a passing muon. 
Searching for correlations between preceding muons and candidate supernova events is the basis of the Super-K ``spallation cut'', which removes $\sim$90\% of spallation events at a 20\% loss in signal efficiency \citep{2016arXiv160607538S}.

While spallation events can be produced anywhere in the tank, radioactive isotopes from the detector itself are much more likely to be located near the detector structure and outside of the standard FV. 
Since most isotope backgrounds are $O(1)$ MeV in energy, restricting the analysis of data outside the FV to higher energies will allow for data from the entire volume to be analyzed.
Indeed, from the analysis in \citet{Mori_phdthesis}, the expected number of background events in 20~s is $4.7\times 10^{-1}$ events above 5~MeV and $4.3\times 10^{-2}$ above 10~MeV inside the FV before application of the spallation cut. 
Application of the spallation cut reduces  $1.6\times 10^{-1}$ events above 5~MeV and $1.8\times 10^{-3}$ events above 10~MeV inside the FV.
Outside of the FV the number is 11 events in 20~s, since the spallation cut has not been extended to this region. 
In conclusion, we expect to be able to carry out a background free analysis inside the FV, and removing events outside of the FV below 10 MeV will allow for purer samples in that region.

\section{Conclusions} \label{sec:concl}
Since the high-energy environment realized within a supernova core is inaccessible to terrestrial experiments, a future observation of supernova neutrino burst is a valuable opportunity for study. However since it is also fraught with uncertainties, a variety of model cases need to be prepared ahead of the next Galactic supernova.
The nuclear EOS is a key ingredient for modeling supernova neutrino light curves. The mass and radius of the neutron star determine the decay timescale of the neutrino luminosity and surface temperature affects the average energy of emitted neutrinos. 
So as to extract the properties of the PNS from a neutrino observation, it is important to understand the impact of the nuclear EOS. 
For this purpose we have estimated the neutrino event rate at Super-Kamiokande based on simulation results of PNS cooling, which are sensitive to the EOS but are not largely influenced by the explosion dynamics. 
In this paper we adopted the four EOS models: the Shen EOS, LS220 EOS, Togashi EOS, and T+S EOS described above.
As already pointed out in \citet{2019ApJ...881..139S}, the neutron star mass is reflected in cumulative event distribution taken as function of backward time from the last observed event. 
Furthermore, in the present study we find that the backward cumulative event distribution is characterized by the EOS. 
In particular, the PNS surface of the Togashi EOS models is rich in heavy nuclei which causes coherent neutrino scattering resulting in sporadic detection of neutrinos over a few minutes in the case of a supernova at 10~kpc. 
Accordingly, the average energy of recoil positrons is expected to be higher since the surface temperature of the PNS is higher compared to the other EOS models.

So as to trace the evolution of average neutrino energy, we propose a new data analysis method for Super-Kamiokande.
With 100 events we can determine the average energy of recoil positrons with an uncertainty of $\Delta \bar E_{e^+} \sim \pm 1$~MeV according to our MC simulations. Thus, the last 100 events are grouped into one bin for evaluating the average energy. Since the neutrino event rate is higher at earlier times bins are chosen so the number in each increases exponentially as counted backward in time. 
This will make it possible to confirm the high average event energy predicted by the Togashi EOS models. 
Note that the event energy is independent of the supernova distance.
Thus, while determining the EOS is challenging this approach provides insight that is complementary to other methods. 
Furthermore, we have introduced a simple expression as in Eq.~(\ref{eq:recstr_enu1st}) for quick calculation of the average neutrino energy. This will be useful for rapid comparison between data and theoretical models when the next Galactic supernova is observed. 

In a forthcoming paper, we are planning to improve our analysis method. 
The detector background, which has been neglected in the present study, will be included in our simulations using a realistic background model based on data from Super-Kamiokande. Although we have assumed that the distance to the supernova is 10 kpc in the present study, the neutrino event rate is proportional to the inverse square of this distance and an uncertainty may be included in the distance measurement. 
We will discuss how to incorporate the distance information into our analysis method considering various types of noise. Finally, we will present a standard procedure for exploring the shape of the supernova neutrino light curve as is already done for the optical light curve.

\begin{acknowledgments}
This work is supported by Grant-in-Aid for Scientific Research (15K05093, 16K17665, 17H02864, 19K03837, 20H00174, 20H01904,  20H01905, 20K03973) and Grant-in-Aid for Scientific Research on Innovative Areas (17H06357, 17H06365, 18H04586, 18H05437, 19H05802, 19H05811, 20H04747) from the Ministry of Education, Culture, Sports, Science and Technology (MEXT), Japan. This work was partially carried out by the joint research program of the Institute for Cosmic Ray Research (ICRR), The University of Tokyo.
For providing high performance computing resources, Research Center for Nuclear Physics in Osaka University, Yukawa Institute of Theoretical Physics in Kyoto University, Center for Computational Astrophysics in National Astronomical Observatory of Japan, Computing Research Center in KEK, JLDG on SINET4 of NII, Information Technology Center in Nagoya University, and Information Technology Center in University of Tokyo are acknowledged.
This work was partly supported by MEXT as ``Program for Promoting Researches on the Supercomputer Fugaku'' (Toward a unified view of the universe: from large scale structures to planets).
\end{acknowledgments}

\bibliography{sample63}{}
\bibliographystyle{aasjournal}



\end{document}